\documentclass[a4paper,11pt]{article}
\usepackage[utf8]{inputenc}
\usepackage{jcappub}

\usepackage{subfig}
\usepackage{epsfig}
\usepackage{graphicx}
\usepackage{color}
\usepackage{amssymb}
\usepackage{amsmath}
\allowdisplaybreaks
\usepackage{mathtools}
\usepackage{url}
\usepackage{natbib}
\usepackage[normalem]{ulem}
\usepackage{afterpage}
\usepackage{float}
\usepackage{ulem}

\newcommand{\id}{\textrm{1\kern-.25em I}}        
\newcommand{\class}{{\sc class}}

\def\dd{{\rm d}}

\newcommand{\Rom}[1]{\MakeUppercase{\romannumeral #1}}

\title{The Hawking Energy in a Perturbed Friedmann-Lema\^{i}tre Universe}
\author[a]{Dennis Stock,}
\author[b]{Enea Di Dio,}
\author[a]{and Ruth Durrer}
\date{\today}

\affiliation[a]{Universit\'e de Gen\`eve, D\'epartement de Physique Th\'eorique and Centre for Astroparticle Physics,
24 quai Ernest-Ansermet, CH-1211 Gen\`eve 4, Switzerland}
\affiliation[b]{Theoretical Physics Department, CERN, 1211 Geneva 23, Switzerland}

\emailAdd{dennis.stock@unige.ch}
\emailAdd{enea.didio@cern.ch}
\emailAdd{ruth.durrer@unige.ch}

\abstract{

Hawking's quasi-local energy definition quantifies the energy enclosed by a spacelike 2-sphere in terms of the amount of lightbending on the sphere caused by the energy distribution inside the sphere. This paper establishes for the first time a direct connection between the formal  mathematical definition of a quasi-local energy and observations, in the  context of cosmological perturbation theory. This is achieved by studying the Hawking Energy of spherical sections of the past lightcone of a cosmic observer in a perturbed Friedmann-Lema\^{i}tre spacetime. We express the Hawking Energy in terms of gauge-invariant perturbation variables and comment on the cosmic observables needed to in principle measure it. We then calculate its angular power spectrum and interpret its contributions. }

\begin{document}

  \begin{minipage}{.45\linewidth}
    \begin{flushleft}
    \end{flushleft}
  \end{minipage}
\begin{minipage}{.45\linewidth}
\begin{flushright}
 {CERN-TH-2023-063}
 \end{flushright}
 \end{minipage}

\maketitle

\section{Introduction}
A general definition of gravitational mass or energy within the theory of general relativity poses a problem and many proposals for such a definition have been brought forward, see for instance \cite{Szabados:2009eka} for a review. In this work, we restrict our analysis to a spacetime describing the evolution of the Universe on large scales governed by general relativity. More specifically, we would like to connect such a definition to cosmological observations, for which the past lightcone of a given cosmic observer is the adequate geometric structure to consider and uniquely defined once a point in spacetime, 'the observer at time $t_0$', is specified. Such a null hypersurface can be described by a generating family of null geodesics in terms of its expansion scalar and shear tensor. Phenomenologically speaking, the presence of energy or mass in a spacetime, either in form of matter or gravitational waves, will affect light propagation. Because the Hawking Energy \cite{Hawking:1968qt} is defined in terms of the expansion scalars of null congruences, it is a natural definition of energy when applied to the lightcone generators, as suggested and studied in \cite{Stock:2020oda} and \cite{Stock:2020drm}. Here, we study the Hawking Energy in the context of linearly perturbed Friedmann-Lema\^itre (FL) spacetimes, providing the standard framework in which cosmological observations are analysed and interpreted, see e.g.~\cite{Durrer:2020fza} for a comprehensive introduction to cosmological perturbation theory.

This work is structured as follows: the Hawking Energy is briefly introduced in Section \ref{sec:Hawkingenergy}, together with its expression in FL spacetimes. In a next step, we derive the expression for the Hawking Energy in spatially flat, linearly perturbed FL spacetimes in longitudinal gauge in Section \ref{sec:perturbedFLRW} and show that it is gauge-invariant. The angular power spectrum of its fluctuations about the background FL spacetime is studied in detail in Section \ref{sec:angularpowerspectra} and we conclude in Section \ref{sec:conclusion}.

In the following, we will use units in which $c=G=1$ and use the metric signature $(-,+,+,+)$. All matter, including a potential cosmological constant, is combined into one energy-momentum tensor, for which we assume the dominant energy condition to hold. The Einstein field equations can then be written as
\begin{equation}
    G_{\mu\nu}=8\pi T_{\mu\nu}\,.
\end{equation}

\section{Hawking Energy and the geometric set-up}
\label{sec:Hawkingenergy}
The Hawking Energy is defined for a spacelike 2-sphere $S$ and aims at measuring the energy enclosed in $S$ by quantifying the amount of lightbending on $S$. Since $S$ is spacelike, there are two distinct null directions orthogonal to $S$, which we will identify with two corresponding null vector fields denoted by $l$ and $n$. These null congruences come with an expansion scalar $\theta_\pm$ and shear tensor $\sigma_{\mu\nu}^\pm$, where $\pm$ refers to $l$ and $n$, respectively. Then, the Hawking Energy of $S$ is defined as \cite{Hawking:1968qt}
\begin{equation}
    E(S) = \frac{\sqrt{A(S)}}{(4\pi)^{3/2}}\left(2\pi+\frac{1}{4}\int_S \theta_+\theta_-\,dS\right)\,,
    \label{eq:EHorig}
\end{equation}
where $A(S)=\int_S\,dS$ is the area of $S$. This definition can be generalised to higher genus surfaces and also to multiple components of closed surfaces, see \cite{Hayward:1993ph} for more details. In the following, we make use of the twice contracted Gauss equation, relating the 2-dimensional Ricci scalar ${}^2R$ of $S$ to the null congruences and to the Ricci tensor of the 4d spacetime via
\begin{equation}
    {}^2R = 16\pi T_{\mu\nu}l^\mu n^\nu+\frac{8\pi}{3}T-\theta_+\theta_-+2\sigma_+^{\mu\nu}\sigma^-_{\mu\nu}\,,
\end{equation}
see Appendix \ref{app:gaussidentity} for a derivation, to rewrite the original definition (\ref{eq:EHorig}) as
\begin{equation}
    E(S) = \frac{\sqrt{A(S)}}{(4\pi)^{3/2}}\int_{S}\left[ 4\pi T_{ln}+\frac{2\pi}{3}T +\frac{1}{2}\sigma_+^{\mu\nu}\sigma^-_{\mu\nu} \right]dS\,,
    \label{eq:E2}
\end{equation}
where we used the Gauss-Bonnet-Theorem for a 2-sphere $S$:
 \begin{equation}
    \int_{S}{}^2R \;dS = 8\pi\,.
\end{equation}

The geometric setup of interest for this work is the past lightcone of a cosmic observer, that is: a point $p$ in spacetime and a timelike, future-pointing vector $u$ at $p$ corresponding to the observer's 4-velocity. The past lightcone $C^-(p)$ of $p$ can then be sliced into a one-parameter family of spacelike 2-surfaces $S_\lambda$, such that $C^-(p)=\cup_\lambda S_\lambda$, with $\lambda\in \mathbb{R}_{\geq0}$ labelling the slices. One can always find a sufficiently small neighbourhood of $p$, so that the lightcone has spherical topology, $C^-(p)~\sim~\mathbb{R}_{\geq0}\times S^2$. But as we move further away from $p$, the lightcone slices $S_\lambda$ can be arbitrarily complicated due to the presence of caustics, triggered for example by inhomogeneities, see e.g. \cite{Stock:2020oda} for more details. But as we are only interested in linear perturbations about an FL background, we can safely assume that all lightcone slices have spherical topology, $S_\lambda\simeq S^2$. In other words, we allow for weak lensing but exclude strong lensing and caustics such that the lightcone can be deformed but its topology always remains unchanged. Of course there are many different choices of foliations to slice the lightcone into spacelike sections, for example constant affine parameter, constant area distance, etc., but because we want to make contact with cosmological observables, we choose a constant redshift slicing in what follows. 

Computing the Hawking Energy for spherical lightcone slices $S_z$, we choose the two null vectors $l$ and $n$ to be past-directed, such that one of them can be identified with the past-directed lightcone generators. For now, we leave the rescaling freedom of $l$ unspecified, but we fix the normalisation of $n$ by demanding $l^\mu n_\mu =-1$. Our final results will be independent of the rescaling choice for $l$.

\subsection{Hawking Energy in FL Spacetimes}
Before discussing the Hawking Energy in the more general context of a linearly perturbed FL setting, we briefly review its expression in FL spacetimes. Evaluating (\ref{eq:E2}) in an FL spacetime, we have $T_{\mu\nu}l^\mu n^\nu=\frac{1}{2}(\rho-P)$, and $T=-\rho+3P$, cf.~(\ref{eq:Tln}) and~(\ref{eq:T}), with $\rho$ the density and $P$ the pressure of the cosmic fluid. We introduce also the, in general  redshift-dependent, equation-of-state parameter $w(z)$ as the ratio of $P(z)$ and $\rho(z)$,
\begin{equation}
    P(z)=w(z)\rho(z)\,.
\label{eq:EOS}
\end{equation}
Furthermore, the lightcone generators in FL spacetimes are shear-free: $\sigma_{\mu\nu}^+=0$. The Hawking Energy then has the simple form
\begin{equation}
    E(z)= E(S_z)= \frac{4\pi}{3}\rho(z) D^3(z)\,,
\label{eq:EFLRW}
\end{equation}
with angular diameter distance\footnote{In general, $D(z)$ is the area distance, which reduces to the angular diameter distance in the spherically symmetric case.} $D$ defined via $A(S_z)=4\pi D^2$.

Here, we are mainly interested in its properties on constant redshift slices as we advance down the lightcone. According to a result by Horowitz and Schmidt \cite{HorowitzSchmidt}, the Hawking Energy is positive for sufficiently small spheres in the neighbourhood of the lightcone vertex point $p$. As it was shown in \cite{Stock:2020drm}, monotonicity of the function $E(z)$ can be established in several spatially flat cases, in particular in the case of dust with a positive cosmological constant, or, more generally, for an equation of state $w(z)\geq 0$.

\section{Hawking Energy in Linearly Perturbed FL}
\label{sec:perturbedFLRW}
In a next step, we include scalar linear perturbations about a spatially flat FL background. For simplicity, we choose longitudinal gauge, our metric convention in conformal time $\eta$ is
\begin{equation}
    ds^2 = a^2(\eta)\left(-(1+2\Psi)d\eta^2+(1-2\Phi)(dr^2+r^2d\Omega)\right)\,,
    \label{eq:metriclong}
\end{equation}
with the Bardeen potentials $\Psi$ and $\Phi$. Note that here and in the following, we normalise the scale factor at the observer $o$ such that $a_o=a(\eta_o)=1$. The components of the energy-momentum tensor in longitudinal gauge are summarised in Appendix \ref{app:energymomentum}. Again, we consider a constant redshift slicing of the observer's past lightcone. Starting from (\ref{eq:E2}), we strictly keep terms only up to linear order in perturbations, hence, the shear term can be neglected because the contraction of the two shear tensors is already of second order:
\begin{equation}
    E(S_z)= \sqrt{\frac{A(S_z)}{4\pi}}\int_{S_z} \epsilon\,dS_z\qquad\text{with}\qquad \epsilon:=T_{\mu\nu}l^\mu n^\nu +\frac{1}{6}T\,.
    \label{eq:EE}
\end{equation}

Now, all quantities are perturbed up to first order.  Note that in general all background quantities, in the following denoted by an overbar, can only depend on redshift and not on direction due to isotropy, whereas all perturbations are functions of redshift and of direction $\hat{n}$ on the observer's celestial sphere. The perturbation of the area element $dS_z$ can either be viewed as perturbation in the solid angle $d\Omega_z$ while keeping the distance $D(z)$ fixed, or as perturbation of the distance for fixed solid angle. Here, for simplicity, we will choose the latter such that $d\Omega_z=d\bar{\Omega}$ and $D=\bar{D}+\delta D$. Summarising, we arrive at the following perturbations:
\begin{align}
    dS_z &= D^2 d\Omega_z\simeq\left(1+2\frac{\delta D}{\Bar{D}}\right)\Bar{D}^2 d\Bar{\Omega}\,,\\
    A &= \int_{S_z}dS_z \simeq \Bar{A}+2\Bar{D}^2\int_{S_z} \frac{\delta D}{\Bar{D}}d\Bar{\Omega}\,,\\
    \epsilon &= \Bar{\epsilon}+ \delta \epsilon\,.
\end{align}
The quantity $\frac{\delta D}{\Bar{D}}$ has been calculated in the literature, see e.g.~\cite{Sasaki:1987ad,Bonvin:2005ps,Yoo:2016vne,Durrer:2020fza}, where it was also shown to be gauge invariant.

Inserting back into (\ref{eq:EE}) yields
\begin{align}
    E(z) &= \sqrt{\frac{\Bar{A}+\delta A}{4\pi}} \int_{S_z} (\bar{\epsilon}+\delta\epsilon)\; (d\bar{S}_z + \delta dS_z)\nonumber\\
    &\simeq \Bar{E}(z)\left(1+\frac{3}{4\pi}\int_{S_z}\frac{\delta D}{\Bar{D}}\,d\Bar{\Omega}\right) +\Bar{D}^3\int_{S_z} \delta\epsilon\, d\Bar{\Omega}\,,
    \label{eq:Eprelim}
\end{align}
where we used $\Bar{\epsilon}=\frac{\bar{\rho}_\mathrm{T}}{3}=\frac{1}{3}(\Bar{\rho}_f+\Lambda)$ with total background density $\bar{\rho}_T$ consisting of a generic cosmic fluid $\bar{\rho}_f$ and a cosmological constant $\Lambda$, see (\ref{eq:epsilonfinal}) and the comment at the beginning of Section \ref{sec:linear epsilon}. Additionally, (\ref{eq:EFLRW}) was used. Note again that all background quantities are functions of redshift only whereas perturbations additionally depend on  direction.
In the next step, we compute the energy-momentum term $\epsilon$ and its perturbation.

\subsection{Computation of $\delta\epsilon$}
In order to compute the perturbations of $\epsilon= T_{ln}+\frac{1}{6}T$, see also (\ref{eq:EE}), we not only have to perturb the energy-momentum tensor, but also both null vectors $l$ and $n$. To simplify the calculation, it is beneficial to adopt coordinates adapted to our lightcone set-up, such as geodesic lightcone coordinates (GLC) introduced in \cite{Gasperini:2011us}. As we want to discuss our final results in longitudinal gauge, we will first construct the null vectors $l$ and $n$ in GLC and then convert back into longitudinal gauge.

\subsubsection{Geodesic Lightcone Coordinates}
\begin{figure}
    \centering
    \includegraphics[scale=0.85]{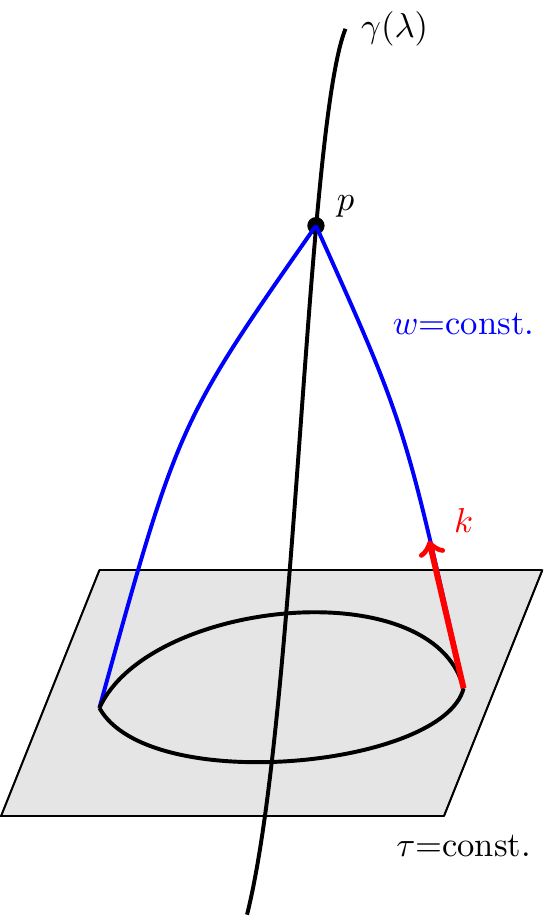}
    \caption{Geodesic Lightcone Coordinates adapted to past lightcones along the observer's worldine $\gamma(\lambda)$. The past lightcone at a point $p$ on this worldline is then given by the hypersurface $w$=const., whereas $\tau$=const. surfaces foliate the lightcone. $k$ is a future-pointing null vector on the lightcone. The union of all $k$ in all directions $(\tilde{\theta}_1,\Tilde{\theta}_2)$ on the celestial sphere at $p$ is the set of light cone generators.}
    \label{fig:GLC}
\end{figure}

Here, we briefly introduce the basic features of GLC and refer the reader to \cite{Gasperini:2011us} and in particular to \cite{Marozzi:2014kua} for a detailed account on GLC. The GLC consist of a timelike coordinate $\tau$, a null coordinate $\mathrm{w}$ and two spacelike angular coordinates $\Tilde{\theta}^a$ with $a=1,2$. Given the observer's worldline, we can decompose spacetime into a 1-parameter family of past lightcones issued from the observer's worldline, the null coordinate $\mathrm{w}$ labels this family of lightcones such that $\mathrm{w}=$const. corresponds to one of these lightcones. The two spacelike angular coordinates $\Tilde{\theta}^a$ specify the direction on the celestial sphere in which a photon travelling on the past lightcone reaches the observer. The timelike coordinate $\tau$ slices the lightcone and can be thought of a coordinate down the lightcone, see also Figure \ref{fig:GLC}. In these coordinates, the line element reads:
\begin{equation}
    ds^2= \Upsilon^2 d\mathrm{w}^2 -2\Upsilon d\mathrm{w}\, d\tau + \gamma_{ab}(d\Tilde{\theta}^a-U^a d\mathrm{w})(d\Tilde{\theta}^b-U^b d\mathrm{w})\,,
    \label{eq:GLClineelement}
\end{equation}
with free functions $\Upsilon$, $U^a$, and $\gamma_{ab}$ the two-dimensional metric of the spacelike surface $\mathrm{w}=$const. and $\tau=$const. In matrix from:
\begin{equation}
    g_{\mu\nu}=
    \begin{pmatrix}
        0 & -\Upsilon & \Vec{0}\\
        -\Upsilon & \Upsilon^2+U^2 & -U_b\\
        \Vec{0}^t & -U_a^t & \gamma_{ab}
    \end{pmatrix}
    \,,~~
    g^{\mu\nu}=
    \begin{pmatrix}
        -1 & -\Upsilon^{-1}&-U^b/\Upsilon\\
        -\Upsilon^{-1}&0&\Vec{0}\\
        -(U^a)^t/\Upsilon&\Vec{0}^t&\gamma^{ab}
    \end{pmatrix}\,,
\end{equation}
where a superscript $^t$ denotes transposition.

The degenerate metric on the past lightcone is then given by $ds^2|_{\mathrm{w}={\rm const.}}$. The past lightcone is generated by all photon null geodesics reaching the observer. The future-pointing null vector associated with each geodesic is given by
\begin{equation}
    k= \frac{\omega}{\Upsilon}\partial_\tau\,,
\end{equation}
with $\omega\in\mathbb{R}$ an arbitrary normalisation constant of $k$. Denoting the observer's 4-velocity by $u$, the expression for the redshift from source to observer in GLC is
\begin{equation}
    1+z = \frac{(u^\mu k_\mu)_s}{(u^\mu k_\mu)_o}= \frac{\Upsilon(\mathrm{w}_o,\tau_o,\Tilde{\theta}^a)}{\Upsilon(\mathrm{w}_o,\tau_s,\Tilde{\theta}^a)}\,.
    \label{eq:redshiftGLC}
\end{equation}
For our purposes, we also need the metric on $S_z$, a constant redshift slice of the lightcone. Fixing the redshift, we can formally invert (\ref{eq:redshiftGLC}) to yield $\tau_s(z)$. With this, we arrive at the metric on $S_z$ by first restricting the general metric (\ref{eq:GLClineelement}) to the lightcone ($\mathrm{w}=$const.) and then further restricting to a constant redshift slice:
\begin{equation}
    ds^2\big|_{S_z} = ds^2\Big|_{
    \begin{subarray}{1}
    \mathrm{w}=\mathrm{const}\\ z=\mathrm{const}
    \end{subarray}}
    = \gamma_{ab}(\mathrm{w}_o,\tau_s(z),\Tilde{\theta}^a)\,d\Tilde{\theta}^ad\Tilde{\theta}^b\,.
\end{equation}
Note that the $\partial_{\Tilde{\theta}^a}$ span $TS_z$.

\subsubsection{Null vectors $l$ and $n$ orthogonal to $S_z$}
We can now move on to explicitly construct the two past-pointing null vectors $l$ and $n$ orthogonal to $S_z$ needed in the definition of the Hawking Energy in GLC by demanding:
\begin{equation}
    l_\mu l^\mu =0\,,\quad n_\mu n^\mu=0\,,\quad l_\mu \left(\partial_{\Tilde{\theta}^a}\right)^\mu=0=n_\mu \left(\partial_{\Tilde{\theta}^a}\right)^\mu\,,\quad l_\mu n^\mu =-1\,.
    \label{eq:normalisation}
\end{equation}
Note that since $g_{\tau\Tilde{\theta}_a}=0$, we can identify $l=-k$, since $l$ should be past-pointing, and $n$ is uniquely determined by the above conditions. We find:
\begin{align}
    l&= -\frac{\omega}{\Upsilon}\partial_\tau \,,\\
    n&= -\frac{1}{\omega}\left(\partial_\mathrm{w}+\frac{1}{2}\Upsilon\partial_\tau+U^a\partial_{\Tilde{\theta}^a}\right)\,.
\end{align}
The conditions (\ref{eq:normalisation}) leave the rescaling of $l$, encoded in the constant $\omega\in\mathbb{R}$, as the only freedom left. Note that our final result does not depend on how $l$ is rescaled, as expected. Next, we convert from GLC into longitudinal gauge following \cite{Marozzi:2014kua}. The full calculation can be found in Appendix \ref{app:lnGLC}, the main result is:
\begin{align}
l&= 
\begin{cases}
l^\eta &= -\frac{\omega}{a^2}(1+\partial_+Q+\partial_-Q-2\Psi)\\
l^r &=\frac{\omega}{a^2}(1+\partial_+Q+3\partial_-Q-2\Psi)\\
l^\theta &= \frac{2\omega}{a^2}\partial_-\vartheta_1\\
l^\varphi &= \frac{2\omega}{a^2}\partial_-\vartheta_2
\end{cases}
\qquad\text{and}\\
n&=
\begin{cases}
n^\eta &= \frac{1}{2\omega}(-1+\partial_+Q+\partial_-Q)\\
n^r &= \frac{1}{2\omega}(-1+\partial_+Q-\partial_-Q)\\
n^\theta &= \frac{1}{\omega} (\partial_+\vartheta_1-U^1)\\
n^\varphi &= \frac{1}{\omega}(\partial_+\vartheta_2-U^2)
\end{cases}\qquad,
\end{align}
where $Q$ and $\vartheta_a$ are defined by (\ref{eq:auxdef}), and $\partial_\pm :=\frac{1}{2}(\partial_\eta \pm \partial_r)$. As a cross-check, it can be verified again that also in GLC, we have $l\cdot n=-1$.

\subsubsection{$\epsilon$ in Linear Order}
\label{sec:linear epsilon}
After constructing $l$ and $n$, we also need to perturb the energy-momentum tensor. The components of the energy-momentum tensor on constant conformal time hypersurfaces can be found in Appendix \ref{app:energymomentum}, leading to the following expressions appearing in $\epsilon$. For convenience and comparison with the literature, we now split the energy momentum tensor into a cosmological constant $\Lambda$, which is unperturbed, and a component due to the presence of the cosmic fluid, denoted by $f$, $T_{\mu\nu}=T^f_{\mu\nu}-\Lambda g_{\mu\nu}$
\begin{align}
    T_{ln}(\eta) &= \Lambda +  T^f_{\mu\nu}l^\mu n^\nu=\Lambda +  \Bar{T}^f_{\mu\nu}(\eta)\left(\Bar{l}^\mu\Bar{n}^\nu+\delta l^\mu \Bar{n}^\mu+\Bar{l}^\mu \delta n^\nu\right) +\delta T^f_{\mu\nu}(\eta)\Bar{l}^\mu\Bar{n}^\nu \nonumber\\
    &= \Lambda +  \frac{\bar\rho_f}{2}(1+\delta)+\frac{\bar{P}_f}{2}\left[-1+2(\Phi+\Psi)-4\partial_-Q-\Pi_L-\Pi_r^r\right]\,,\label{eq:Tln}\\
    T(\eta) &= T_\mu^\mu = -4\Lambda  -\bar{\rho}_f(1+\delta)+3\Bar{P}_f(1+\Pi_L)\,,\label{eq:T}
\end{align}
where we used
\begin{align}
    \Bar{T}_{\mu\nu}(\eta)\Bar{l}^\mu\Bar{n}^\nu &= \Lambda + \frac{1}{2}(\Bar{\rho}_f-\Bar{P}_f)\\
    \Bar{T}^{f}_{\mu\nu}(\eta)\delta l^\mu \Bar{n}^\mu &= \frac{\Bar{\rho}_f}{2}(\partial_+Q+\partial_-Q-2\Psi)-\frac{\Bar{P}_f}{2}(\partial_+Q+3\partial_-Q-2\Psi)\\
    \Bar{T}^{f}_{\mu\nu}(\eta) \Bar{l}^\mu \delta n^\nu &= -\frac{\Bar{\rho}_f}{2}(\partial_+Q+\partial_-Q)+\frac{\Bar{P}_f}{2}(\partial_+Q-\partial_-Q)\\
    \delta T^{f}_{\mu\nu}(\eta)\Bar{l}^\mu\Bar{n}^\nu &= \frac{\Bar{\rho}_f}{2}(\delta+2\Psi)-\frac{\Bar{P}_f}{2}(\Pi_L+\Pi^r_r-2\Phi)
    \,.
\end{align}
Upon using $\partial_-Q=\frac{1}{2}(\Phi+\Psi)$, we thus find
\begin{equation}
    \epsilon(\eta)= \frac{\bar{\rho}_f}{3}(1+\delta)+\frac{\Lambda}{3}-\frac{1}{2}\bar{P}_f\,\Pi_r^r\,.
\end{equation}
Next, we need to convert from conformal time to observed redshift as required by (\ref{eq:Eprelim}): $\epsilon(\eta)\rightarrow\epsilon(z)$ . Recalling that conformal time is related to the background redshift $\eta=\eta(\bar{z})$, we find to first order:
\begin{equation}
    \epsilon(z) = \bar{\epsilon}(\bar{z})+\delta\epsilon(z)= \bar{\epsilon}(z-\delta z)+\delta\epsilon(z)=\bar{\epsilon}(z)-\frac{d\bar{\epsilon}}{d\bar{z}}\delta z+\delta\epsilon(z) = \epsilon(\eta)-\frac{d\bar{\epsilon}}{d\bar{z}}\delta z\,,
\end{equation}
with $\bar{\epsilon}=\frac{1}{3}(\bar{\rho}_f+\Lambda)$. The derivative term converts a background quantity from constant time to constant redshift slices. This correction is only relevant for background quantities, for a first order term $g$ we can set $g(\eta)=g(z)$ to first order because its correction term is of second order. The continuity equation
\begin{equation}
    \partial_\eta \bar{\rho}_f = -3(\bar{\rho}_f+\bar{P}_f)\mathcal{H}\,,
\end{equation}
for the background cosmic fluid with equation of state $\bar{P}_f=\bar{w}\, \bar{\rho}_f$, cf. (\ref{eq:EOS}), can be rewritten in terms of redshift using $\partial_{\bar{z}}=-\frac{1}{H}\partial_\eta$:
\begin{equation}
    \frac{d\bar{\rho}_f}{d\bar{z}}=3\bar{\rho}_f\frac{1+\bar{w}}{1+\bar{z}}\,,
\end{equation}
where $H$ denotes the Hubble rate and $\mathcal{H}=H a$ the conformal Hubble rate. Hence, we have 
\begin{equation}
    \epsilon(z) = \bar{\rho}_f\left\{\frac{1}{3}(1+\delta)-\frac{\bar{w}}{2}\Pi^r_r-\frac{1+\bar{w}}{1+\bar{z}}\delta z\right\} +\frac{\Lambda}{3}\,.
    \label{eq:epsilonfinal}
\end{equation}

\subsection{Hawking Energy in Linear Order}
Putting everything back into the expression for the Hawking Energy (\ref{eq:Eprelim}), we have (in longitudinal gauge)
\begin{equation}
    E(z) = \bar{E}(z)\left\{1+\frac{1}{4\pi}\int\,d\bar{\Omega}\left[3\frac{\delta D}{\bar{D}}+\frac{\bar{\rho}_{f}}{\bar{\rho}_{f}+\Lambda}\left(\delta-3(1+\bar{w})\frac{\delta z}{1+\bar{z}}-\frac{3}{2}\bar{w}\,\Pi^r_r\right)\right]\right\}\,.
    \label{eq:Elong}
\end{equation}
The area distance and redshift fluctuations have already been computed for instance in \cite{Scaccabarozzi:2018vux}, and read:
\begin{align}
    \frac{\delta z}{1+\bar{z}} &= \mathcal{H}_o \delta\eta_o +\left[\Psi+\bar{n}_iV^i\right]^o_s -\int_0^{r(z)}\! dr'\,(\dot{\Psi}+\dot{\Phi})\qquad\text{and}\label{eq:deltaz}\\
    \frac{\delta D}{\bar{D}} &= \left(1-\frac{1}{\mathcal{H}r}\right)\left[(\mathcal{H}\delta\eta)_o-[V^i\bar{n}_i+\Psi]_o^s-\int_0^{r(z)}\! dr'(\dot{\Phi}+\dot{\Psi})\right]\nonumber\\
    &-\Phi+\frac{\delta\eta_o}{r}-(\bar{n}_iV^i)_o+\frac{1}{r(z)}\int_0^{r(z)}\!dr'\left(1-\frac{r(z)-r'}{2r'}\Delta_\Omega\right)(\Phi+\Psi)\,,\label{eq:deltaD}
\end{align}
with $\delta\eta_o := -\int_0^{\eta_o} d\eta\, a\Psi$. 
In the above expression, the integrals are performed from the observer position at $r=0$ to $r(z)$ along the unperturbed photon geodesic, $\eta = \eta_0-r', x_i=r'\bar n_i$, $\Delta_\Omega$ denotes the angular Laplacian and $V$ is a velocity potential such that the velocity in longitudinal gauge is given by $V_i = -\partial_i V$. An index $_o$ denotes evaluation at the observer position while and index $]^s_o$ denotes evaluation at the 'source' position (i.e. at $(x_s^\mu)=(\eta_0-r(z),r(z)\bar n_i)$) minus evaluation at the observer position. For a pressureless matter fluid, $w=0$, the time lapse  becomes, 
\begin{equation}
\delta\eta_o = -\int_0^{\eta_o} d\eta\, a\Psi=-V_o\,.
\end{equation}
Note also that the above expressions include all observer terms. These cannot be neglected in our case, since the fluctuations of the Hawking Energy correspond to the monopole of the 'Hawking Energy surface density', see (\ref{eq:Elong}), where the integrand denotes the Hawking Energy surface density. 

In order to verify that the monopole fluctuation of the Hawking Energy is finite, we relate the density fluctuation in longitudinal gauge $\delta\equiv\delta_\mathrm{long}$ to the density fluctuation in comoving gauge $\mathcal{D}$ via
\begin{equation}
    \mathcal{D} = \delta_\mathrm{long}+3(1+\bar{w})\mathcal{H}V\,,
\end{equation}
with velocity potential $V$. This then leads to the final result
\begin{align}
    E(z) &= \bar{E}(z)\left\{1+\frac{1}{4\pi}\int d\bar{\Omega}\; \mathcal{E}\right\}\qquad\text{with}\\
    \mathcal{E}&:=3\frac{\delta D}{\bar{D}}+\frac{\bar{\rho}_f}{\bar{\rho}_f+\Lambda}\left(\mathcal{D}-3(1+\bar{w})\left(\mathcal{H}V+\frac{\delta z}{1+\bar{z}}\right)-\frac{3}{2}\bar{w}\,\Pi^r_r\right) \,.
    \label{eq:Efinal}
\end{align}
$\mathcal{E}$ can be interpreted as the fluctuation in the surface energy density on $S_z$. This expression is gauge-invariant and its monopole is indeed finite, because each individual term has a finite monopole once the anisotropic stress term is neglected, as we shall do in the following. For the term $\mathcal{H}V+\frac{\delta z}{1+\bar{z}}$, this has been shown explicitly in \cite{Castorina:2021xzs}.

\subsection{Relation to Cosmic Observables}
Before moving on to the discussion of the angular power spectrum, we would like to establish a more direct connection to cosmic observables. It turns out that all terms in (\ref{eq:Efinal}) can be related to quantities which are observed in practice. All background quantities can be inferred from background cosmological parameters, for instance determined by the Planck satellite \cite{Planck}.

Regarding the fluctuation quantities, we first consider  the relative fluctuation in the area distance $\frac{\delta D}{\bar{D}}$. It is best measured 'statistically' with the baryon acoustic oscillations at different redshifts,  see, e.g.~\cite{Anderson:2012sa,DESI:2023bgx}. Also  luminosity distance $D_L$ is measured with supernova type Ia observations, see, e.g. ~\cite{Brout:2022vxf}. Both distances are related via the well-known Etherington relation
\begin{equation}
    D_L=(1+z)^2\,D\quad.
\end{equation}
Using the Etherington relation for fixed observed redshift $z$, we find that the relative fluctuations of luminosity and area distance are equal:
\begin{equation}
    \frac{\delta D}{\bar{D}} = \frac{\delta D_L}{\bar{D}_L}\quad.
\end{equation}
Therefore measurements of the area distance or the luminosity distance both can provide $ \frac{\delta D}{\bar{D}}$. Let us also consider the density fluctuation at constant redshift $\delta_z(\Vec{n},z)$ which has already been calculated in \cite{Bonvin_2011}:
\begin{equation} \label{eq:density_z}
    \delta_z(\Vec{n},z)= \frac{\rho(\Vec{n},z)-\bar{\rho}(z)}{\bar{\rho}(z)}= \frac{\delta\rho}{\bar{\rho}} - \frac{d\bar{\rho}}{d\bar{z}}\frac{\delta z}{\bar{\rho}}=\delta-3(1+\bar{w})\frac{\delta z}{1+\bar{z}} = \mathcal{D} -3(1+\bar{w})\left(\frac{\delta z}{1+\bar{z}}+\mathcal{H}V\right)\quad.
\end{equation}
This quantity can in principle be measured since the Bardeen potentials are measurable\footnote{ In principle, only second derivatives of the gravitational potential are physical and measurable. However, even if eq.~\eqref{eq:density_z} appears at first sight to depend on the gravitational potential, the final observable is not sensitive to  constant and gradient modes of the gravitational potential, as discussed in details in section~\ref{sec:monopole}.} (their sum via weak lensing and $\Psi$ alone via its effect on accelerating massive particles, see~\cite{Tutusaus:2022cab} for  detailed suggestions).  Also the velocity is measurable via redshift space distortions, see, e.g.~\cite{Lange:2021zre}.
Finally, the density contrast in comoving gauge is related to the Bardeen potential $\Phi$ via the Poisson equation, see~\cite{Durrer:2020fza}.

In principle these observations would have to be performed over the full sky to obtain the total Hawking Energy, which is not feasible with present techniques as we cannot reliably count galaxies in regions obscured by the Milky Way. However, with present and near future surveys (DESI, LSST, Euclid and SKA \cite{DESI:2016fyo,DESI:2023bgx,Abate:2012za,Abell:2009aa,Amendola:2016saw,Laureijs:2011gra,Maartens:2015mra}),
we shall cover substantial fractions of sky, which will allow us to infer the number count angular power spectrum for harmonics 
$\ell \gtrsim 5$.

In the standard model the anisotropic stress can be neglected, but if it is present it can be inferred observationally via the difference of the measured Bardeen potentials using Einstein’s equations, see~\cite{Tutusaus:2022cab} for details.

Note that in the present derivation we have used Einstein’s equation in several steps, so our results are in principle only valid within GR. On the other hand, the gravitational equations of arbitrary modifications of GR can be written in the form of Einstein’s equation with a strange energy momentum tensor $T_{\mu\nu} = T^{\rm matter}_{\mu\nu}  + T^{\rm mod}_{\mu\nu} $. The second term, the modification of the energy momentum tensor, can contain anisotropic stresses, have phantom behavior etc. But it is clear that each term in 
\begin{equation}
    \mathcal{E}(z,\vec{n})=
    3\frac{\delta D_L}{\bar{D}_L}+\frac{\bar{\rho}_f}{\bar{\rho}_f+\Lambda}\, \delta_z -\frac{3}{2}\bar{w}\,\Pi^r_r
    \label{eq:E2contrib}
\end{equation} 
is in principle measurable as we expect it from a gauge invariant cosmological perturbation variable.

\section{Angular Power Spectra}
\label{sec:angularpowerspectra}
Having obtained the expression (\ref{eq:Efinal}) for the Hawking Energy in a linearly perturbed FL spacetime, we move on to study its angular power spectrum to analyse its dominant contributions and its redshift dependence. Since the background energy, $\bar{E}$,  is determined by cosmological parameters, we study its relative fluctuation, $\delta E$, given by
\begin{eqnarray}
    \delta E &:=&\frac{E-\bar{E}}{\bar{E}} = \frac{1}{4\pi}\int\,d\bar{\Omega}\;\mathcal{E}(\bar{n},z)\qquad\text{with}\label{eq:deltaE}\\ 
    \mathcal{E} &:=& 3\frac{\delta D}{\bar{D}}+\frac{\bar{\rho}_f}{\bar{\rho}_f+\Lambda}\left[\mathcal{D}-3(1+\Bar{w})\left(\mathcal{H}V+\frac{\delta z}{1+\bar{z}}\right)\right]\,,
    \label{eq:epsilon}
\end{eqnarray}
after neglecting anisotropic stress. According to (\ref{eq:deltaE}), the fluctuation $\delta E$ of the Hawking Energy itself is a monopole, however the integrand $\mathcal{E}$, the surface density of the Hawking Energy, also has higher multipoles. Before computing the fluctuation of the Hawking Energy in Section \ref{sec:monopole}, we briefly study the angular power spectrum of its surface density $\mathcal{E}$ for the standard $\Lambda$CDM model. To this aim, we modify the angular transfer functions (see Appendix~\ref{app:CLASS} for more details) of the \class{} code \cite{CLASS, DiDio:2013bqa, DiDio:2016ykq} with the underlying cosmological parameters based on the most recent Planck data \cite{Planck} given in Table \ref{tab:cosmicparams}.

Of course we cannot perform a true ensemble average in cosmology. But we make, as usual, an 'ergodic hypothesis' assuming that the mean square fluctuations on a given angular scale $2\pi/\ell$ is given by the average over many patches of size $2\pi/\ell$. This quantity is actually  determined by the power spectrum. More precisely it is $\ell(\ell+1)C_\ell/2\pi$. For large $\ell$ this is a good approximation while for small $\ell$ it is not accurate, see~\cite{Durrer:2020fza} for more details on this problem of cosmic variance.

\begin{figure}
    \centering
    \includegraphics[scale=0.7]{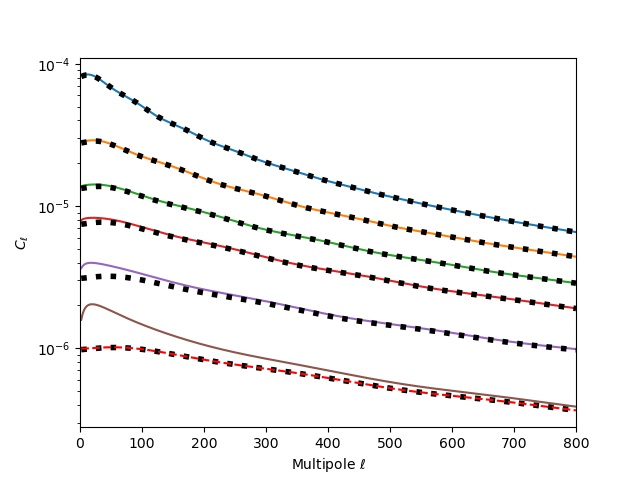}
    \caption{Angular power spectrum $C_\ell^\mathcal{E}$ at redshifts $z= 0.5,~ 1,~ 1.5,~ 2,~ 3,~ 5$ from top to bottom up to $\ell=800$. The black dotted curves correspond to the respective density contribution only, see (\ref{eq:Efinal}). The red dashed curve for the highest redshift $z=5$ corresponds to all terms except lensing. Hence, the difference in the amplitudes between the density contribution only and all terms is due to lensing cross correlations.}
    \label{fig:Cle}
\end{figure}

\begin{table}[]
    \centering
    \begin{tabular}{c|ccccc} 
       Parameter & $h$& $\Omega_bh^2$ & $\Omega_ch^2$&  $n_s$&$\log(10^{10}A_s)$\\
       \hline
       Value & $0.6732$ & $0.022383$ & $0.12011$ & 0.96605 & 3.0448\\
    \end{tabular}
    \caption{Cosmological parameters based on Planck data~\cite{Planck} used in the computation of the power spectra in this section.}
    \label{tab:cosmicparams}
\end{table}
Figure \ref{fig:Cle} shows the angular power spectra $C_\ell^\mathcal{E}$ for different redshifts up to $\ell_\mathrm{max}=800$. Of all the terms appearing in (\ref{eq:Efinal}), the density $\mathcal{D}$ is clearly the dominant contribution across the studied range of $\ell$ for lower redshifts. The overall power drops with increasing redshift as well as with increasing $\ell$. As density fluctuations dominate the spectrum, the growth in density fluctuations over time due to cosmic structure formation explains the redshift dependence. For larger redshifts, however, also other contributions become relevant, especially on large scales, the lensing term integrated along the line-of-sight in (\ref{eq:deltaD}).

\subsection{Monopole} 
\label{sec:monopole}
\begin{figure}[h]
    \centering
    \includegraphics[scale=0.7]{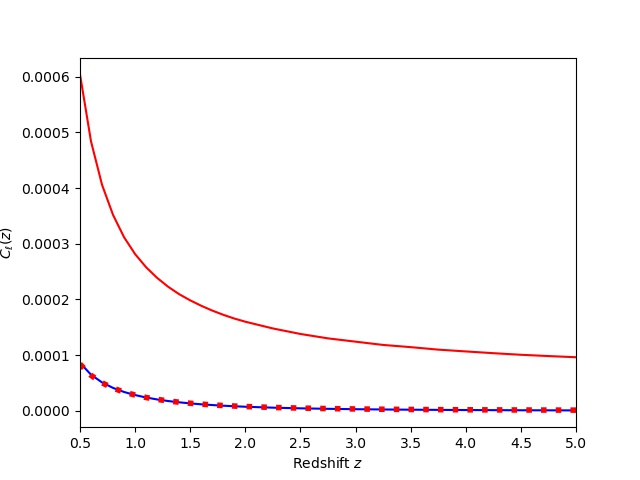}
    \caption{Monopole (blue) and dipole (red) as function of redshift. The dotted red curve is the dipole without the peculiar velocity of the observer. Thus, the dominant part of the dipole amplitude is due to the observer's peculiar velocity.}
    \label{fig:monopoledipole}
\end{figure}

The expectation value for $\delta E$ given in  (\ref{eq:deltaE}) vanishes of course within linear perturbation theory about an FL Universe, but its variance is given by the monopole of the power spectrum. More precisely, in a typical Universe we expect the Hawking energy to deviate by about $\sqrt{4\pi C_0}\bar{E}$ from the background value. Of course, for the monopole, cosmic variance is large, for Gaussian fluctuations it amounts to  $\sigma_0=\sqrt{2}$, see~\cite{Durrer:2020fza}. Figure \ref{fig:monopoledipole} shows the monopole and dipole contributions of $\mathcal{E}$ as functions of redshift $z$. Firstly, we note that the dipole amplitude is significantly larger than the monopole or any of the higher multipoles, compare Figure \ref{fig:Cle}. It turns out that the dominant contribution to the dipole is the observer's peculiar velocity\footnote{ The observer velocity is assumed to be described by the linear velocity field. In principle our observer velocity can be measured in the CMB rest-frame~\cite{Kogut:1993ag,Lineweaver:1996xa,WMAP:2008ydk,Planck:2013kqc,Planck,Planck:2020qil}. However the difference is expected to be within the cosmic variance~\cite{Mitsou:2019ocs,Desjacques:2020zue}.}. If the peculiar velocity terms at the observer are removed, the dipole is of the same order of magnitude as the other multipoles. In general, the monopole amplitude is decreasing with redshift, i.e.~the energy fluctuations $\delta E$ become smaller as redshift increases. Again this is expected, because $\delta E$ is dominated mainly by density fluctuations, 
see~Figure~\ref{fig:Cle}.

\subsubsection*{Numerical IR-Cancellations}
It turns out that special care is needed when computing the monopole, because some of the terms appearing in (\ref{eq:epsilon}) are IR-divergent\footnote{The calculation of the angular power spectrum  contains integrals over wavenumber $k$ some of which, in the case of the monopole $\ell=0$, diverge for $k\rightarrow 0$.}, despite their total sum being finite. In the following we present a way to ensure that all IR-divergencies cancel out explicitly, so that the numerical expression for the monopole in (\ref{eq:deltaE}) is  independent of the numerical IR cutoff.
As we can see from eqs.~(\ref{eq:deltaz}-\ref{eq:deltaD}) the Hawking Energy perturbation depends on the gravitational potential. However we know that a constant gravitational potential and its gradient are not physical in GR and, therefore, they can not impact our observable. In absence of anisotropic stress, this has been already shown for all the individual terms appearing in eqs.~\eqref{eq:Efinal}, namely the luminosity distance fluctuation $\delta D/\bar D$ and $ \left( \mathcal{H}V+\frac{\delta z}{1+\bar{z}} \right)$.

In order to accurately evaluate the monopole of the angular power spectrum, we will make use of an IR cancellation explicitly, such that our prediction is insensitive to the numerical IR cutoff. This approach has been worked out in detail in Ref.~\cite{Castorina:2021xzs} for the galaxy power spectrum. Here we adapt it to the Hawking Energy perturbation by first computing the full 3-dimensional correlation function $\xi \left( r_1, r_2, \mu= \mathbf{r}_1 \cdot \mathbf{r}_2 \right)$ and then to relate it to the angular power spectrum through
\begin{equation}
\label{eq:xitocl}
      C_\ell \left( r_1 , r_2 \right) = 2 \pi \int d \mu \ \xi \left( r_1,r_2,\mu \right) \mathcal{L}_\ell \left( \mu \right)  
\end{equation}
which for the monopole simply reduces to
\begin{equation}
      C_0 \left( r_1 , r_2 \right) = 2 \pi \int d \mu \ \xi \left( r_1,r_2,\mu \right) \, .
\end{equation}
Following Refs.~\cite{Tansella:2018sld,Castorina:2021xzs} we write the correlation function in the form
\begin{eqnarray} \label{eq:full_correlation}
&& \hspace{-0.8cm}\xi \left( r_1 , r_2 , \mu = \hat {\mathbf{r}}_1 \cdot \hat {\mathbf{r}}_2 \right)
\nonumber \\
&=& \sum_{i,n} J_{ni}^{(r)} \left(r_1 , r_2 , \mu \right) D_1 \left( r_1 \right) D_1 \left( r_2 \right) I^n_i \left( r \right) 
\nonumber \\
&&+
\sum_{i,n} J_{ni}^{(r_1)} \left(r_1 ,r_2,{ \mu} \right) D_1 \left( r_1 \right)  I^n_i \left( r_1 \right) +
\sum_{i,n} J_{ni}^{(r_2)} \left(r_2 ,r_1 ,{ \mu} \right) D_1 \left( r_2 \right)  I^n_i \left( r_2 \right) 
\nonumber \\
&& +
 D_1 \left( r_2 \right) \int_0^{r_1}\dd \chi_1 \sum_{i,n} J_{ni}^{( \Delta \chi_1 )} \left(r_1 , r_2 , \mu,\chi_1 \right) I^n_i \left( \Delta \chi_1  \right) 
\nonumber \\
&& +
 D_1 \left( r_1 \right)  \int_0^{r_2}\dd \chi_2 \sum_{i,n} J_{ni}^{(\Delta \chi_2)} \left(r_1 , r_2  , \mu,\chi_2 \right) I^n_i \left( \Delta \chi_2 \right) 
 \nonumber \\
&& +
 \int_0^{r_1}\dd \chi_1 \sum_{i,n} J_{ni}^{(\chi_1 )} \left(r_1 , \mu,\chi_1 \right) I^n_i \left( \chi_1  \right) 
+
  \int_0^{r_2}\dd \chi_2 \sum_{i,n} J_{ni}^{(\chi_2)} \left(  r_2  , \mu,\chi_2 \right) I^n_i \left( \chi_2 \right) 
\nonumber \\
&&
+
\int_0^{r_1}\dd \chi_1  \int_0^{r_2} \dd \chi_2 \sum_{i,n} J_{ni}^{(\chi)} \left(r_1 , r_2 , \mu,\chi_1,\chi_2 \right)  I^n_i \left( \chi \right) 
\nonumber \\
&&+
J_{\sigma_2}\left( r_1 , r_2, \mu \right) \sigma_2   + J_{\sigma4} \left( r_1 , r_2, \mu \right) \sigma_4 \, ,
\end{eqnarray}
where $D_1$ denotes the linear density growth factor normalized to $1$ today. We have also introduced the generalized Hankel transform the matter power spectrum today,
\begin{equation}
    I^n_\ell \left( s \right) =  \int \frac{dq}{2 \pi^2} q^2 P \left( q \right) \frac{j_\ell \left( q s \right)}{\left( q s\right)^n} \, ,
\end{equation}
and 
\begin{equation}
\sigma_i = \int \frac{\dd q }{2 \pi^2} q^{2-i} P \left( q \right)  \, .
\end{equation}
The arguments of the generalized Hankel transform for integrated terms are of forms
\begin{eqnarray}
\Delta \chi_2 &=& \sqrt{r_1^2+\chi_2^2 -2 r_1 \chi_2 \mu}
\\
\Delta \chi_1  &=&\sqrt{\chi_1^2+r_2^2 -2 \chi_1 r_2 \mu}
\\
\chi&=&\sqrt{\chi_1^2+\chi_2^2 -2 \chi_1 \chi_2 \mu}
\end{eqnarray}
and all the coefficients $J^{(a)}_{ni}$ are defined in appendix~\ref{app:coeff}.
The IR cancellation allows us to replace
\begin{equation}
            I^4_0 \left( s \right) \rightarrow \tilde I^4_0 \left( s \right) =  \int \frac{dq}{2 \pi^2} q^2 P \left( q \right) \frac{j_0 \left( q s \right) -1}{\left( q s\right)^4}
\end{equation}
and insured the IR convergence of the integral without relying on numerical cancellations.

\section{Conclusion}
\label{sec:conclusion}
Amongst all quasi-local energy definitions within general relativity, the Hawking Energy gets an intuitive interpretation as the energy of the Universe accessible to a cosmic observer, once it is applied to sections of the observer's past lightcone. Being computed on the geometric structure related to observations, one might hope to directly relate it to cosmic observables. In this paper, we show that this can indeed be achieved within an important class of cosmological spacetimes, namely linearly perturbed Friedmann spacetimes. Because the result (\ref{eq:Efinal}) is gauge-invariant, it must be related to observables. It turns out that there are two main contributions on top of its background value, recall (\ref{eq:E2contrib}): fluctuations in the area or luminosity distance, and matter density fluctuations on constant redshift slices. Both can in principle be measured, provided observations are sufficiently detailed and accurate. In a numerical evaluation we have found that the result at low redshift is determined by the density fluctuations in comoving gauge while at higher redshift also the lensing convergence term contributes appreciably.

\acknowledgments
We are grateful to Matthias Bartelmann and Giuseppe Fanizza for valuable discussions.
This work is supported by the Swiss National Science Foundation.

\vspace{2cm}

\appendix

\begin{center}{\LARGE\bf Appendix}\end{center}


\section{Twice-contracted Gauss identity}
\label{app:gaussidentity}
Given our spacelike surface $S$ and the two distinct null vectors $l$ and $n$ orthogonal to $S$ and satisfying the conditions (\ref{eq:normalisation}), the induced metric $q_{\mu\nu}$ on $S$ is given by
\begin{equation}
    q_{\mu\nu} = g_{\mu\nu} +l_\mu n_\nu+n_\mu l_\nu
\end{equation}
together with the associated projection operator $q_\mu^\nu$ onto $S$:
\begin{equation}
    q_\mu^\nu = \delta_\mu^\nu+l_\mu n^\nu+n_\mu l^\nu\,.
\end{equation}
Introducing the covariant derivative ${}^2D$ on $S$, the two-dimensional Ricci tensor ${}^2R_{\mu\nu}$ associated with ${}^2D$ on $S$ is given by
\begin{equation}
    {}^2D_\mu {}^2D_\alpha v^\mu -{}^2D_\alpha {}^2D_\mu v^\mu = {}^2R_{\alpha\mu} v^\mu\,,
\end{equation}
with $v\in TS$ arbitrary. The deformation tensors of $S$ along $l$ and $n$
\begin{equation}
    \Theta_{\mu\nu}:= q_\mu^\alpha q_\nu^\beta\, \nabla_\alpha l_\beta\qquad,\qquad \Xi_{\mu\nu}:= q_\mu^\alpha q_\nu^\beta\, \nabla_\alpha n_\beta
\end{equation}
can be split into a trace part proportional to the respective expansion scalar $\theta_\pm$ and traceless part given by the respective shear tensor $\sigma^\pm_{\mu\nu}$ \footnote{Note that photon 4-velocities are gradients of the phase of the wave vector so that their deformation tensor is symmetric.},
\begin{equation}
    \Theta_{\mu\nu} = \frac{1}{2}\theta_+ q_{\mu\nu} +\sigma^+_{\mu\nu}\qquad,\qquad \Xi_{\mu\nu} = \frac{1}{2}\theta_- q_{\mu\nu} +\sigma^-_{\mu\nu}\,.
\end{equation}
As it was shown in \cite{Gourgoulhon:2005ng}, cf. (6.38), we can then express the Ricci tensor and scalar on $S$ as 
\begin{align}
    {}^2R_{\alpha\beta} &= q_\alpha^\mu q_\beta^\nu q_\sigma^\rho \,R^\sigma_{\nu\rho\mu} -\theta_+\Xi_{\alpha\beta} -\theta_-\Theta_{\alpha\beta} + \Theta_{\alpha\mu}\Xi^\mu_\beta + \Xi_{\alpha\mu}\Theta^\mu_\beta\,,\\
    {}^2R &= q^{\mu\nu}q^{\rho\alpha}R_{\alpha\nu\rho\mu} -\theta_+\theta_- +2\sigma^+_{\alpha\beta}\sigma_-^{\alpha\beta}\,.\label{eq:2dricci}
\end{align}
Next, we decompose the Riemann tensor into its Weyl and Ricci parts, see for example \cite{Wald:1984rg}:
\begin{equation}
    R_{\alpha\beta\gamma\delta} = C_{\alpha\beta\gamma\delta} +\underbrace{g_{\alpha[\gamma}R_{\delta]\beta} -g_{\beta[\gamma}R_{\delta]\alpha}}_{=:\,\Rom{1}} -\underbrace{\frac{R}{3} g_{\alpha[\gamma}g_{\delta]\beta}}_{=:\,\Rom{2}}\,.
\end{equation}
Explicit calculation yields
\begin{align}
    \Rom{1}\cdot q^{\alpha\gamma} q^{\beta\delta}= R+2R_{ln}\,,\\
    \Rom{2} \cdot q^{\alpha\gamma} q^{\beta\delta}= \frac{R}{3}\,,\\
    C_{\alpha\beta\gamma\delta}q^{\alpha\gamma} q^{\beta\delta} = 2C_{lnnl}\,.\label{eq:Weylterm}
\end{align}
Contracting the shear evolution equation for $\sigma^+_{\mu\nu}$,
\begin{equation}
    \nabla_l\sigma^+_{\mu\nu} = -\theta_+ \sigma^+_{\mu\nu} -C_{l\mu\nu l}\,,
\end{equation}
with $n^\mu n^\nu$, we see that the Weyl tensor term (\ref{eq:Weylterm}) is vanishing because $n\perp \sigma^+$:
\begin{equation}
    C_{lnnl} = -\theta_+\sigma^+_{\mu\nu}n^\mu n^\nu -n^\mu n^\nu \nabla_l \sigma^+_{\mu\nu}=0\,,
\end{equation}
since
\begin{equation}
    0=\nabla_l(\sigma^+_{\mu\nu} n^\mu n^\nu)=  \underbrace{\sigma^+_{\mu\nu}n^\mu}_{=0} \nabla_l n^\nu+ \underbrace{\sigma^+_{\mu\nu}n^\nu}_{=0} \nabla_l n^\mu +n^\mu n^\nu \nabla_l \sigma^+_{\mu\nu}\,.
\end{equation}
Upon using the Einstein field equations $R_{\mu\nu}-\frac{1}{2}Rg_{\mu\nu}= 8\pi T_{\mu\nu}$ and the above results yields for (\ref{eq:2dricci}):
\begin{equation}
    {}^2R = 16\pi T_{ln}+\frac{8\pi}{3}T-\theta_+\theta_-+2\sigma_{\mu\nu}^+\sigma_-^{\mu\nu}\,.
\end{equation}

\section{Conversion of $l$ \& $n$ from GLC to longitudinal gauge}
\label{app:lnGLC}
Recall that the null vectors $l$ and $n$ in GLC are given by
\begin{align}
    l&= -\frac{\omega}{\Upsilon}\partial_\tau \,,\\
    n&= -\frac{1}{\omega}\left(\partial_\mathrm{w}+\frac{1}{2}\Upsilon\partial_\tau+U^a\partial_{\Tilde{\theta}^a}\right)\,.
\end{align}
In order to determine their expressions in longitudinal gauge, we need to perform a basis change from GLC $y^\mu=(\tau,\mathrm{w},\Tilde{\theta}^a)$ to longitudinal gauge $x^\mu=(\eta,r,\theta_a)$ (with $\theta_1=\theta$, $\theta_2=\varphi$), such that for example
\begin{equation}
    l=l^\mu(y)\frac{\partial}{\partial y^\mu} = \underbrace{l^\mu(y(x))\frac{\partial x^\nu}{\partial y^\mu}}_{=l^\nu(x)}\frac{\partial}{\partial x^\nu}\,.
    \label{eq:coordtrafo}
\end{equation}
Thus, we need the relation between GLC and longitudinal coordinates $y(x)$ as well as its Jacobian in order to compute its inverse $\frac{\partial x^\nu}{\partial y^\mu}$ . The  coordinate change from GLC to longitudinal coordinates has already been studied in \cite{Marozzi:2014kua}, here we will adopt the notation of this paper and restate the findings of the author. The only difference between the results summarised here and \cite{Marozzi:2014kua} is due to the different definition of the perturbed metric in longitudinal gauge, cf. (\ref{eq:metriclong}), i.e. $\Psi\leftrightarrow\Phi$ to convert between the two notations. Using the auxiliary definitions
\begin{align}
    \eta_\pm &=\eta\pm r\,,\quad \partial_\pm=\frac{\partial}{\partial \eta_\pm}=\frac{1}{2}(\partial_\eta \pm\partial_r)\,,\nonumber\\
    P(\eta,r,\theta_a)&=\int_{\eta_\mathrm{in}}^\eta d\eta' \,\frac{a(\eta')}{a(\eta)}\Psi(\eta',r,\theta_a)\,,\nonumber\\
    Q(\eta_+,\eta_-,\theta_a)&= \frac{1}{2}\int_{\eta_o}^{\eta_-}dx\,(\Phi+\Psi)(\eta_+,x,\theta_a)\,,\nonumber\\
    \vartheta^a(\eta_+,\eta_-,\theta^a) &=\frac{1}{2}\int_{\eta_o}^{\eta_-}dx\, [\gamma^{ab}_0 \partial_bQ](\eta_+,x,\theta_a)\quad\text{with}\quad \gamma_{ab}^0=\mathrm{diag}(r^2,r^2\sin^2\theta)\,,
    \label{eq:auxdef}
\end{align}
we have:
\begin{align}
    \tau &= \int_{\eta_\mathrm{in}}^\eta d\eta'a(\eta') +a\,P\quad\text{with}\nonumber\\
    \frac{d\tau}{d\eta}&= a(1+\Psi)\,,\quad \frac{d\tau}{dr}=a\,\partial_r P\,,\quad \frac{d\tau}{d\theta_a}=a\,\partial_{\theta_a} P\,,\\
    \mathrm{w} &= \eta_++Q\quad\text{with}\nonumber\\
    \frac{d\mathrm{w}}{d\eta}&= 1+\partial_+Q+\partial_-Q\,,\quad \frac{d\mathrm{w}}{dr}=1+\partial_+Q-\partial_-Q\,,\quad \frac{d\mathrm{w}}{d\theta_a}=\partial_{\theta_a}Q\,,\\
    \Tilde{\theta}^a &=\theta^a + \vartheta^a(\eta_+,\eta_-,\theta_a)\quad\text{with}\nonumber\\
    \frac{d\Tilde{\theta}^a}{d\eta} &= \partial_+\vartheta^a+\partial_-\vartheta^a\,,\quad \frac{d\Tilde{\theta}^a}{dr} = \partial_+\vartheta^a -\partial_-\vartheta^a\,,\quad \frac{d\Tilde{\theta}^a}{d\theta^b} = \delta^a_b+\partial_b\vartheta^a\,.
\end{align}
Calculating the components of the inverse Jacobian yields:
\begin{align}
    \frac{d\eta}{d\tau}&=\frac{1}{a}(1+\partial_rP-\Psi)\,,\quad \frac{d\eta}{d\mathrm{w}}=-\partial_rP\,,\quad \frac{d\eta}{d\Tilde{\theta}^a}=-\partial_{\theta^a}P\,,\nonumber\\
    \frac{dr}{d\tau}&= \frac{1}{a}(-1-\partial_rP+\Psi-2\partial_-Q)\,,\quad \frac{dr}{d\mathrm{w}}= 1+\partial_rP-\partial_+Q+\partial_-Q\,,\nonumber\\
    \frac{dr}{d\Tilde{\theta}^a}&= \partial_{\theta^a}(P-Q)\,,\nonumber\\
    \frac{d\theta^b}{d\tau}&= -\frac{2}{a}\partial_-\vartheta^b\,,\quad \frac{d\theta^b}{d\mathrm{w}}= (-\partial_++\partial_-)\vartheta^b\,,\quad \frac{d\theta^b}{d\Tilde{\theta}^a}= \delta^b_a-\partial_{\theta^a}\vartheta^b\,.
\end{align}
In order to calculate the components of $l$ and $n$ in longitudinal gauge via (\ref{eq:coordtrafo}), we also need the first order expansions of $\Upsilon$ and $U^a$:
\begin{align}
    \Upsilon^{-1}&=-g^{w\tau}\simeq \frac{1}{a}(1+\partial_+Q+\partial_-Q-\Psi-\partial_rP)\\
    \Upsilon&= a(1-\partial_+Q-\partial_-Q+\Psi+\partial_rP)\\
    U^a&=\partial_+\vartheta^a+\partial_-\vartheta^a\frac{1}{r^2}\partial_\theta P\;\delta^a_\theta-\frac{1}{r^2\sin^2\theta}\partial_\varphi P\;\delta_\varphi^a\,.
\end{align}
Using (\ref{eq:coordtrafo}), the vectors $l$ and $n$ in longitudinal gauge are given by
\begin{align}
l&= 
\begin{cases}
l^\eta &= -\frac{\omega}{a^2}(1+\partial_+Q+\partial_-Q-2\Psi)\\
l^r &=\frac{\omega}{a^2}(1+\partial_+Q+3\partial_-Q-2\Psi)\\
l^\theta &= \frac{2\omega}{a^2}\partial_-\vartheta_1\\
l^\varphi &= \frac{2\omega}{a^2}\partial_-\vartheta_2
\end{cases}
\qquad\text{and}\\
n&=
\begin{cases}
n^\eta &= \frac{1}{2\omega}(-1+\partial_+Q+\partial_-Q)\\
n^r &= \frac{1}{2\omega}(-1+\partial_+Q-\partial_-Q)\\
n^\theta &= \frac{1}{\omega} (\partial_+\vartheta_1-U^1)\\
n^\varphi &= \frac{1}{\omega}(\partial_+\vartheta_2-U^2)
\end{cases}\qquad.
\end{align}

\section{Energy-momentum tensor in perturbed FL}
\label{app:energymomentum}
In the following, $T_{\mu\nu}$ always denotes the matter energy momentum tensor, without cosmological constant, i.e. it corresponds to $T^f_{\mu\nu}$ in section \ref{sec:linear epsilon}. Following \cite{Durrer:2020fza}, the components of the energy-momentum tensor in a linearly perturbed FL spacetime in longitudinal gauge are given by
\begin{align}
    T^\eta_\eta &=-\Bar{\rho}(1+\delta)\,,\quad T^\eta_j=(\Bar{\rho}+\Bar{P})v_j\,,\quad T^j_\eta=-(\Bar{\rho}+\Bar{P})v^j\,,\nonumber \\
    T^i_j &= \Bar{P}\left[(1+\Pi_L)\delta^i_j+\Pi^i_j\right]\,,
\end{align}
with $\mathrm{tr}\,\Pi=\Pi^i_i=0$. Hence
\begin{align}
    T_{\eta\eta}&= a^2\Bar{\rho}(1+\delta+2\Psi)\\
    T_{\eta r}&= -a^2(\Bar{\rho}+\Bar{P})v^r\\
    T_{\eta \theta}&= -a^2r^2(\Bar{\rho}+\Bar{P})v^\theta\\
    T_{\eta \varphi}&= -a^2r^2\sin^2\theta(\Bar{\rho}+\Bar{P})v^\varphi\\
    T_{rr}&= a^2\Bar{P}(1+\Pi_L+\Pi_r^r-2\Phi)\\
    T_{r\theta}&= a^2r^2\Bar{P}\,\Pi^\theta_r\\
    T_{r\varphi}&= a^2r^2\sin^2\theta\Bar{P}\,\Pi^\varphi_r\\
    T_{\theta\theta}&= a^2r^2\Bar{P}(1+\Pi_L+\Pi_\theta^\theta-2\Phi)\\
    T_{\theta\varphi}&= a^2r^2\sin^2\theta \Bar{P}\,\Pi_\theta^\varphi\\
    T_{\varphi\varphi}&= a^2r^2\sin^2\theta\Bar{P}(1+\Pi_L+\Pi_\varphi^\varphi-2\Phi)\,.
\end{align}

\section{Correlation function coefficients}
\label{app:coeff}
Here we summarize the non-vanishing coefficients of the 3-dimensional correlation functions expressed in terms of generalized Hankel transform of the matter power spectrum through eq.~\eqref{eq:full_correlation}, for matter fluctuations only, i.e.~$w=0$. These coefficients are also summarised in the attached Mathematica file.
\begin{align}
    J_{00}^{(r)} 
&=    f_2 \left\{-\frac{1}{10} \mathcal{R}_1 \mathcal{R}_2 \mathcal{H}_2     (r_2-\mu  r_1) \left(3 a_1 \mathcal{H}_0^2 \Omega_{M0}   \left(r_1^2-2 \mu  r_1 r_2+r_2^2\right)-10\right)
    \right.
    \nonumber \\
    & \qquad \left.
    +\frac{\mathcal{R}_1   (\mathcal{H}_2 r_2-1) (r_2-\mu  r_1) \left(3 a_1 \mathcal{H}_0^2 \Omega_{M0}   \left(r_1^2-2 \mu  r_1 r_2+r_2^2\right)-10\right)}{10 r_2}
      \right.
    \nonumber \\
    & \qquad \left.
    -\frac{3 a_1 \mathcal{R}_2 \mathcal{H}_0^2 \mathcal{H}_2 \Omega_{M0}     (2 \mathcal{H}_1 r_1-1) (\mu  r_1-r_2) \left(r_1^2-2 \mu  r_1 r_2+r_2^2\right)}{10 \mathcal{H}_1 r_1}
      \right.
    \nonumber \\
    & \qquad \left.
    +\frac{3 a_1 \mathcal{H}_0^2 \Omega_{M0}   (2 \mathcal{H}_1 r_1-1) (\mathcal{H}_2 r_2-1) (\mu  r_1-r_2) \left(r_1^2-2 \mu  r_1 r_2+r_2^2\right)}{10 \mathcal{H}_1 r_1 r_2}\right\}
    \nonumber \\
    &
    +f_1 \left\{-\frac{1}{10} \mathcal{R}_1 \mathcal{R}_2 \mathcal{H}_1     (r_1-\mu  r_2) \left(3 a_2 \mathcal{H}_0^2 \Omega_{M0}   \left(r_1^2-2 \mu  r_1 r_2+r_2^2\right)-10\right)
      \right.
    \nonumber \\
    & \qquad \left.
    -\frac{3 a_2 \mathcal{R}_1 \mathcal{H}_0^2 \mathcal{H}_1 \Omega_{M0}     (2 \mathcal{H}_2 r_2-1) (\mu  r_2-r_1) \left(r_1^2-2 \mu  r_1 r_2+r_2^2\right)}{10 \mathcal{H}_2 r_2}
      \right.
    \nonumber \\
    & \qquad \left.
    +\frac{\mathcal{R}_2   (\mathcal{H}_1 r_1-1) (r_1-\mu  r_2) \left(3 a_2 \mathcal{H}_0^2 \Omega_{M0}   \left(r_1^2-2 \mu  r_1 r_2+r_2^2\right)-10\right)}{10 r_1}
      \right.
    \nonumber \\
    & \qquad \left.
    -\frac{3 a_2 \mathcal{H}_0^2 \Omega_{M0}   (\mathcal{H}_1 r_1-1) (2 \mathcal{H}_2 r_2-1) (r_1-\mu  r_2) \left(r_1^2-2 \mu  r_1 r_2+r_2^2\right)}{10 \mathcal{H}_2 r_1 r_2}\right\}
    \nonumber \\
    &
    +f_1 f_2 \left\{
    \frac{1}{5} \mathcal{R}_1 \mathcal{R}_2 \mathcal{H}_1 \mathcal{H}_2       
      \right.
    \nonumber \\
    & \qquad \left.
    \times 
    \left[-\mu  \left(r_1^2 (\mathcal{H}_1 r_1+2)+r_2^2 (3 \mathcal{H}_1 r_1+2)\right)
      \right. \right.
    \nonumber \\
    & \qquad \left. \left.
    +\mu ^2 r_1 r_2 (2 \mathcal{H}_1 r_1+1)+r_1 r_2 (\mathcal{H}_1 r_1+3)+\mathcal{H}_1 r_2^3
    \right. \right. 
    \nonumber \\
    & \qquad \left. \left.
    +\mathcal{H}_2 (r_1-\mu  r_2) \left(r_1^2-2 \mu  r_1 r_2+r_2^2\right)\right]
      \right.
    \nonumber \\
    & \qquad \left.
    -\frac{\mathcal{R}_1 \mathcal{H}_1     (\mathcal{H}_2 r_2-1)}{5 r_2} 
    \left[-\mu  \left(r_1^2 (\mathcal{H}_1 r_1+2)+r_2^2 (3 \mathcal{H}_1 r_1+2)\right)
             \right. \right.
    \nonumber \\
    & \qquad \quad \left. \left. 
    +\mu ^2 r_1 r_2 (2 \mathcal{H}_1 r_1+1)+r_1 r_2 (\mathcal{H}_1 r_1+3)+\mathcal{H}_1 r_2^3\right]
      \right.
    \nonumber \\
    & \qquad \left.
    -\frac{\mathcal{R}_2 \mathcal{H}_2     (\mathcal{H}_1 r_1-1)}{5 r_1}
          \right.
    \nonumber \\
    & \qquad \left. \times
    \left[\mathcal{H}_2 (r_1-\mu  r_2) \left(r_1^2-2 \mu  r_1 r_2+r_2^2\right)-2 \mu  \left(r_1^2+r_2^2\right)+\mu ^2 r_1 r_2+3 r_1 r_2\right]
      \right.
    \nonumber \\
    & \qquad \left.
    -\frac{  (\mathcal{H}_1 r_1-1) (\mathcal{H}_2 r_2-1) \left(2 \mu  r_1^2-\left(\mu ^2+3\right) r_1 r_2+2 \mu  r_2^2\right)}{5 r_1 r_2}\right\}
    +\mathcal{R}_1 \mathcal{R}_2  \, ,
    \\
    J_{02}^{(r)} 
&=
    f_2 \left\{-\frac{1}{7} \mathcal{R}_1 \mathcal{R}_2 \mathcal{H}_2    (r_2-\mu  r_1) \left(3 a_1\mathcal{H}_0^2 \Omega_{M0}   \left(r_1^2-2 \mu  r_1 r_2+r_2^2\right)-7\right)
    \right.
    \nonumber \\
& \qquad \left.
    +\frac{\mathcal{R}_1  (\mathcal{H}_2 r_2-1) (r_2-\mu  r_1) \left(3 a_1\mathcal{H}_0^2 \Omega_{M0}   \left(r_1^2-2 \mu  r_1 r_2+r_2^2\right)-7\right)}{7 r_2}
        \right.
    \nonumber \\
& \qquad \left.
    -\frac{3 a_1\mathcal{R}_2 \mathcal{H}_0^2 \mathcal{H}_2 \Omega_{M0}    (2 \mathcal{H}_1 r_1-1) (\mu  r_1-r_2) \left(r_1^2-2 \mu  r_1 r_2+r_2^2\right)}{7 \mathcal{H}_1 r_1}
        \right.
    \nonumber \\
& \qquad \left.
    +\frac{3 a_1\mathcal{H}_0^2 \Omega_{M0}  (2 \mathcal{H}_1 r_1-1) (\mathcal{H}_2 r_2-1) (\mu  r_1-r_2) \left(r_1^2-2 \mu  r_1 r_2+r_2^2\right)}{7 \mathcal{H}_1 r_1 r_2}\right\}
    \nonumber \\
    &
    +f_1 \left\{-\frac{1}{7} \mathcal{R}_1 \mathcal{R}_2 \mathcal{H}_1    (r_1-\mu  r_2) \left(3 a_2\mathcal{H}_0^2 \Omega_{M0}   \left(r_1^2-2 \mu  r_1 r_2+r_2^2\right)-7\right)
        \right.
    \nonumber \\
& \qquad \left.
    -\frac{3 a_2\mathcal{R}_1 \mathcal{H}_0^2 \mathcal{H}_1 \Omega_{M0}    (2 \mathcal{H}_2 r_2-1) (\mu  r_2-r_1) \left(r_1^2-2 \mu  r_1 r_2+r_2^2\right)}{7 \mathcal{H}_2 r_2}
        \right.
    \nonumber \\
& \qquad \left.
    +\frac{\mathcal{R}_2  (\mathcal{H}_1 r_1-1) (r_1-\mu  r_2) \left(3 a_2\mathcal{H}_0^2 \Omega_{M0}   \left(r_1^2-2 \mu  r_1 r_2+r_2^2\right)-7\right)}{7 r_1}
        \right.
    \nonumber \\
& \qquad \left.
    -\frac{3 a_2\mathcal{H}_0^2 \Omega_{M0}  (\mathcal{H}_1 r_1-1) (2 \mathcal{H}_2 r_2-1) (r_1-\mu  r_2) \left(r_1^2-2 \mu  r_1 r_2+r_2^2\right)}{7 \mathcal{H}_2 r_1 r_2}\right\}
    \nonumber \\
    &
    +f_1 f_2 \left\{\frac{2}{7} \mathcal{R}_1 \mathcal{R}_2 \mathcal{H}_1 \mathcal{H}_2      \left[-\mu  \left(r_1^2 (\mathcal{H}_1 r_1+2)+r_2^2 (3 \mathcal{H}_1 r_1+2)\right)
        \right. \right.
    \nonumber \\
& \qquad \left. \left.
    +\mu ^2 r_1 r_2 (2 \mathcal{H}_1 r_1+1)+r_1 r_2 (\mathcal{H}_1 r_1+3)+\mathcal{H}_1 r_2^3
           \right. \right.
    \nonumber \\
& \qquad \left. \left. 
    +\mathcal{H}_2 (r_1-\mu  r_2) \left(r_1^2-2 \mu  r_1 r_2+r_2^2\right)\right]
        \right.
    \nonumber \\
& \qquad \left.
    -\frac{2 \mathcal{R}_1 \mathcal{H}_1}{7 r_2}    (\mathcal{H}_2 r_2-1) \left[-\mu  \left(r_1^2 (\mathcal{H}_1 r_1+2)+r_2^2 (3 \mathcal{H}_1 r_1+2)\right)
         \right. \right.
    \nonumber \\
& \qquad \qquad \left. \left.   
    +\mu ^2 r_1 r_2 (2 \mathcal{H}_1 r_1+1)+r_1 r_2 (\mathcal{H}_1 r_1+3)+\mathcal{H}_1 r_2^3 \right]
        \right.
    \nonumber \\
& \qquad \left.
    -\frac{2 \mathcal{R}_2 \mathcal{H}_2}{7 r_1}
       (\mathcal{H}_1 r_1-1) \left[\mathcal{H}_2 (r_1-\mu  r_2) \left(r_1^2-2 \mu  r_1 r_2+r_2^2\right)
             \right. \right.
    \nonumber \\
& \qquad \qquad \left. \left.  
    -2 \mu  \left(r_1^2+r_2^2\right)+\mu ^2 r_1 r_2+3 r_1 r_2 \right]
        \right.
    \nonumber \\
& \qquad \left.
    -\frac{2  (\mathcal{H}_1 r_1-1) (\mathcal{H}_2 r_2-1) \left(2 \mu  r_1^2-\left(\mu ^2+3\right) r_1 r_2+2 \mu  r_2^2\right)}{7 r_1 r_2}\right\} \, ,
        \\
    J_{04}^{(r)} 
&=
    f_2 \left\{-\frac{9}{70} a_1\mathcal{R}_1 \mathcal{R}_2 \mathcal{H}_0^2 \mathcal{H}_2 \Omega_{M0}      (r_2-\mu  r_1) \left(r_1^2-2 \mu  r_1 r_2+r_2^2\right)
    \right.
    \nonumber \\
    & \left. \qquad
    +\frac{9 a_1\mathcal{R}_1 \mathcal{H}_0^2 \Omega_{M0}    (\mathcal{H}_2 r_2-1) (r_2-\mu  r_1) \left(r_1^2-2 \mu  r_1 r_2+r_2^2\right)}{70 r_2}
        \right.
    \nonumber \\
    & \left. \qquad
    -\frac{9 a_1\mathcal{R}_2 \mathcal{H}_0^2 \mathcal{H}_2 \Omega_{M0}    (2 \mathcal{H}_1 r_1-1) (\mu  r_1-r_2) \left(r_1^2-2 \mu  r_1 r_2+r_2^2\right)}{70 \mathcal{H}_1 r_1}
        \right.
    \nonumber \\
    & \left. \qquad
    +\frac{9 a_1\mathcal{H}_0^2 \Omega_{M0}  (2 \mathcal{H}_1 r_1-1) (\mathcal{H}_2 r_2-1) (\mu  r_1-r_2) \left(r_1^2-2 \mu  r_1 r_2+r_2^2\right)}{70 \mathcal{H}_1 r_1 r_2}\right\}
    \nonumber \\
    &
    +f_1 \left\{-\frac{1}{70} 9 a_2\mathcal{R}_1 \mathcal{R}_2 \mathcal{H}_0^2 \mathcal{H}_1 \Omega_{M0}      (r_1-\mu  r_2) \left(r_1^2-2 \mu  r_1 r_2+r_2^2\right)
        \right.
    \nonumber \\
    & \left. \qquad
    -\frac{9 a_2\mathcal{R}_1 \mathcal{H}_0^2 \mathcal{H}_1 \Omega_{M0}    (2 \mathcal{H}_2 r_2-1) (\mu  r_2-r_1) \left(r_1^2-2 \mu  r_1 r_2+r_2^2\right)}{70 \mathcal{H}_2 r_2}
        \right.
    \nonumber \\
    & \left. \qquad
    +\frac{9 a_2\mathcal{R}_2 \mathcal{H}_0^2 \Omega_{M0}    (\mathcal{H}_1 r_1-1) (r_1-\mu  r_2) \left(r_1^2-2 \mu  r_1 r_2+r_2^2\right)}{70 r_1}
        \right.
    \nonumber \\
    & \left. \qquad
    -\frac{9 a_2\mathcal{H}_0^2 \Omega_{M0}  (\mathcal{H}_1 r_1-1) (2 \mathcal{H}_2 r_2-1) (r_1-\mu  r_2) \left(r_1^2-2 \mu  r_1 r_2+r_2^2\right)}{70 \mathcal{H}_2 r_1 r_2}\right\}
    \nonumber \\
    &
    +f_1 f_2 \left\{\frac{3}{35} \mathcal{R}_1 \mathcal{R}_2 \mathcal{H}_1 \mathcal{H}_2      \left[-\mu  \left(r_1^2 (\mathcal{H}_1 r_1+2)+r_2^2 (3 \mathcal{H}_1 r_1+2)\right)
        \right.\right.
    \nonumber \\
    & \left. \left. \qquad
    +\mu ^2 r_1 r_2 (2 \mathcal{H}_1 r_1+1)+r_1 r_2 (\mathcal{H}_1 r_1+3)
    +\mathcal{H}_1 r_2^3
            \right.\right.
    \nonumber \\
    & \left. \left. \qquad
    +\mathcal{H}_2 (r_1-\mu  r_2) \left(r_1^2-2 \mu  r_1 r_2+r_2^2\right)\right]
            \right.
    \nonumber \\
    & \left. \qquad
    -\frac{3 \mathcal{R}_1 \mathcal{H}_1}{35 r_2}    (\mathcal{H}_2 r_2-1) \left[-\mu  \left(r_1^2 (\mathcal{H}_1 r_1+2)+r_2^2 (3 \mathcal{H}_1 r_1+2)\right)
            \right.\right.
    \nonumber \\
    & \left. \left. \qquad
    +\mu ^2 r_1 r_2 (2 \mathcal{H}_1 r_1+1)+r_1 r_2 (\mathcal{H}_1 r_1+3)+\mathcal{H}_1 r_2^3 \right]
\right.
    \nonumber \\
    & \left. \qquad
    -\frac{3 \mathcal{R}_2 \mathcal{H}_2}{35 r_1}     (\mathcal{H}_1 r_1-1) \left[\mathcal{H}_2 (r_1-\mu  r_2) \left(r_1^2-2 \mu  r_1 r_2+r_2^2\right)
           \right.            \right.
    \nonumber \\
    & \left. \left. \qquad
    -2 \mu  \left(r_1^2+r_2^2\right)+\mu ^2 r_1 r_2+3 r_1 r_2 \right]
                    \right.
    \nonumber \\
    & \left. \qquad
    -\frac{3  (\mathcal{H}_1 r_1-1) (\mathcal{H}_2 r_2-1) \left(2 \mu  r_1^2-\left(\mu ^2+3\right) r_1 r_2+2 \mu  r_2^2\right)}{35 r_1 r_2}\right\} \, ,
            \\
    J_{20}^{(r)}
&= 
    -\frac{9 \mathcal{H}_0^2 \Omega_{M0} }{2 \mathcal{H}_1 \mathcal{H}_2 r_1 r_2} \left\{ \left(r_1^2-2 \mu  r_1 r_2+r_2^2\right) (a_1\mathcal{R}_2 \mathcal{H}_2 r_2 (\mathcal{H}_1 r_1 (\mathcal{R}_1-2)+1)
    \right. 
\nonumber \\
& \left. \qquad
    +a_2\mathcal{R}_1 \mathcal{H}_1 r_1 (\mathcal{H}_2 r_2 (\mathcal{R}_2-2)+1)) \right\}
    \nonumber \\
    &
    +\frac{3 f_2}{2 \mathcal{H}_1 r_1 r_2}  \left(r_1^2-2 \mu  r_1 r_2+r_2^2\right)
    \nonumber \\
& \qquad
    \left\{3 a_1\mathcal{H}_0^2 \Omega_{M0} (\mu  r_1-r_2) (\mathcal{H}_1 r_1 (\mathcal{R}_1-2)+1) (\mathcal{H}_2 r_2 (\mathcal{R}_2-1)+1)
    \right. \nonumber \\ & \left. \qquad
    +2 \mathcal{R}_1 \mathcal{R}_2 \mathcal{H}_1 \mathcal{H}_2^2 r_1 r_2  \right\} 
    \nonumber \\
    &
    -\frac{3 f_1}{2 \mathcal{H}_2 r_1 r_2}  \left(r_1^2-2 \mu  r_1 r_2+r_2^2\right)
    \nonumber \\ & \qquad 
    \left\{3 a_2\mathcal{H}_0^2 \Omega_{M0} (r_1-\mu  r_2) (\mathcal{H}_1 r_1 (\mathcal{R}_1-1)+1) (\mathcal{H}_2 r_2 (\mathcal{R}_2-2)+1)
        \right. \nonumber \\ & \left. \qquad
    -2 \mathcal{R}_1 \mathcal{R}_2 \mathcal{H}_1^2 \mathcal{H}_2 r_1 r_2  \right\}
    \nonumber \\
    &
    -\frac{3 f_1 f_2 }{r_1 r_2}    \left(r_1^2-2 \mu  r_1 r_2+r_2^2\right) 
        \nonumber \\ & \qquad 
    \left\{(\mathcal{H}_1 r_1-1) (\mathcal{R}_2 \mathcal{H}_2 r_2   (\mathcal{H}_2 (r_1-\mu  r_2)+\mu )-\mathcal{H}_2 \mu  r_2+\mu )
         \right. \nonumber \\ & \left. \qquad
    -\mathcal{R}_1 \mathcal{H}_1 r_1   (\mathcal{H}_1 (r_2-\mu  r_1) (\mathcal{H}_2 r_2 (\mathcal{R}_2-1)+1)
             \right. \nonumber \\ & \left. \qquad
    +\mathcal{R}_2 \mathcal{H}_2 r_2   (\mathcal{H}_2 r_1-\mathcal{H}_2 \mu  r_2+\mu )-\mathcal{H}_2 \mu  r_2+\mu ) \right\} \, ,
    \\
    J_{40}^{(r)}
&=  
    \frac{81 a_1a_2\mathcal{H}_0^4 \Omega_{M0}^2  \left(r_1^2-2 \mu  r_1 r_2+r_2^2\right)^2 (\mathcal{H}_1 r_1 (\mathcal{R}_1-2)+1) (\mathcal{H}_2 r_2 (\mathcal{R}_2-2)+1)}{4 \mathcal{H}_1 \mathcal{H}_2 r_1 r_2}
    \nonumber \\
    &
    -\frac{27 a_1\mathcal{R}_2 f_2 \mathcal{H}_0^2 \mathcal{H}_2^2 \Omega_{M0}    \left(r_1^2-2 \mu  r_1 r_2+r_2^2\right)^2 (\mathcal{H}_1 r_1  (\mathcal{R}_1-2)+1)}{2 \mathcal{H}_1 r_1}
    \nonumber \\
    &
    -\frac{27 a_2\mathcal{R}_1 f_1 \mathcal{H}_0^2 \mathcal{H}_1^2 \Omega_{M0}    \left(r_1^2-2 \mu  r_1 r_2+r_2^2\right)^2 (\mathcal{H}_2 r_2 (\mathcal{R}_2-2)+1)}{2 \mathcal{H}_2 r_2}
    \nonumber \\
    &
    +9 \mathcal{R}_1 \mathcal{R}_2 f_1 f_2 \mathcal{H}_1^2 \mathcal{H}_2^2      \left(r_1^2-2 \mu  r_1 r_2+r_2^2\right)^2
      \, ,
    \\
    J_{40}^{(r_1)} &=    
    \frac{27 a_1 \mathcal{H}_0^3 \Omega_{M0} r_1^3 }{4 \mathcal{H}_1 \mathcal{H}_2 r_2} \left\{ (\mathcal{H}_1 r_1  (\mathcal{R}_1-2)+1) (2 f_0 (\mathcal{H}_0 \mathcal{H}_2 r_2 (\mathcal{R}_2-1)+\mathcal{H}_0-\mathcal{H}_2)
    \right. \nonumber \\
    & \left. \qquad
    -3 \mathcal{H}_0 \Omega_{M0} (\mathcal{H}_2 r_2 (\mathcal{R}_2-1)+1)) \right\}
    \nonumber \\
    &
    +\frac{9 \mathcal{R}_1 f_1 \mathcal{H}_0 \mathcal{H}_1^2 r_1^4}{2 \mathcal{H}_2 r_2}
    \left\{
        (3 \mathcal{H}_0 \Omega_{M0} (\mathcal{H}_2 r_2 (\mathcal{R}_2-1)+1)
        \right. \nonumber \\
    & \left. \qquad
    -2 f_0 (\mathcal{H}_0 \mathcal{H}_2 r_2 (\mathcal{R}_2-1)+\mathcal{H}_0-\mathcal{H}_2)) \right\} \, ,
       \\
    J_{31}^{(r_1)} &=    
    -\frac{9 \mathcal{H}_0 r_1^2}{2 \mathcal{H}_1 \mathcal{H}_2 r_2}
    \left\{\left(\mathcal{R}_1 \mathcal{H}_1 (f_0 (\mathcal{H}_0 (3 a_1 \mathcal{H}_0 \mu  \Omega_{M0} r_1   (\mathcal{R}_2 \mathcal{H}_2 r_2  +1)
       \right.      \right. \nonumber \\
    & \left. \left. \qquad \quad
    +2 \mathcal{H}_2 r_2 (\mathcal{R}_2-1)+2)-2 \mathcal{H}_2)-3 \mathcal{H}_0 \Omega_{M0} (\mathcal{H}_2 r_2 (\mathcal{R}_2-1)+1))
           \right.     \right. \nonumber \\
    & \left. \left. \qquad \quad
    -3 a_1 f_0 \mathcal{H}_0^2 \mu  \Omega_{M0} (2 \mathcal{H}_1 r_1-1) (\mathcal{R}_2 \mathcal{H}_2 r_2  +1)\right) \right\}
        \nonumber \\
    &
    -\frac{9 f_1 \mathcal{H}_0 r_1 }{2 \mathcal{H}_2 r_2}
    \left\{(r_1 (\mathcal{H}_1 r_1 (\mathcal{R}_1-1)+1) (2 f_0 (\mathcal{H}_0 \mathcal{H}_2 r_2 (\mathcal{R}_2-1)+\mathcal{H}_0-\mathcal{H}_2)
                    \right. \nonumber \\
    & \left. \qquad
    -3 \mathcal{H}_0 \Omega_{M0} (\mathcal{H}_2 r_2 (\mathcal{R}_2-1)+1))
                    \right. \nonumber \\
    & \left. \qquad
    -2 f_0 \mu  (\mathcal{H}_1 r_1-1) (\mathcal{R}_1 \mathcal{H}_1 r_1  +1) (\mathcal{R}_2 \mathcal{H}_2 r_2  +1)) \right\} \, ,
          \\
    J_{11}^{(r_1)} &=   
    \frac{9 f_0 f_1 \mathcal{H}_0 \mu  r_1   (\mathcal{H}_1 r_1 (\mathcal{R}_1-1)+1) (\mathcal{R}_2 \mathcal{H}_2 r_2  +1)}{5 \mathcal{H}_2 r_2}
            \nonumber \\
    &
    +\frac{3 \mathcal{R}_1 \mathcal{H}_0 r_1 }{10 \mathcal{H}_2 r_2}
    \left\{(2 f_0 (r_1 (\mathcal{H}_0 \mathcal{H}_2 r_2 (\mathcal{R}_2-1)+\mathcal{H}_0-\mathcal{H}_2)+5 \mu  (\mathcal{R}_2 \mathcal{H}_2 r_2  +1))
                        \right. \nonumber \\
    & \left. \qquad
    -3 \mathcal{H}_0 \Omega_{M0} r_1 (\mathcal{H}_2 r_2 (\mathcal{R}_2-1)+1)) \right\} \, ,
              \\
    J_{13}^{(r_1)} &=   \frac{9 f_0 f_1 \mathcal{H}_0 \mu  r_1   (\mathcal{H}_1 r_1 (\mathcal{R}_1-1)+1) (\mathcal{R}_2 \mathcal{H}_2 r_2  +1)}{5 \mathcal{H}_2 r_2}
                \nonumber \\
    &
    +\frac{3 \mathcal{R}_1 \mathcal{H}_0 r_1^2 }{10 \mathcal{H}_2 r_2}
    \left\{(2 f_0 (\mathcal{H}_0 \mathcal{H}_2 r_2 (\mathcal{R}_2-1)+\mathcal{H}_0-\mathcal{H}_2)
                            \right. \nonumber \\
    & \left. \qquad
    -3 \mathcal{H}_0 \Omega_{M0} (\mathcal{H}_2 r_2 (\mathcal{R}_2-1)+1)) \right\} \, ,
                  \\
    J_{\sigma_2} &=  \frac{3 f_0^2 \mathcal{H}_0^2 \mu  (\mathcal{R}_1 \mathcal{H}_1 r_1  +1) (\mathcal{R}_2 \mathcal{H}_2 r_2  +1)}{\mathcal{H}_1 \mathcal{H}_2 r_1 r_2} \, ,
                  \\
    J_{\sigma_4} &= 
    \frac{9 \mathcal{H}_0^2}{4 \mathcal{H}_1 \mathcal{H}_2 r_1 r_2}
    \left\{(2 f_0 (\mathcal{H}_0 \mathcal{H}_1 r_1 (\mathcal{R}_1-1)+\mathcal{H}_0-\mathcal{H}_1)
       \right. \nonumber \\
    & \left. \qquad
    -3 \mathcal{H}_0 \Omega_{M0} (\mathcal{H}_1 r_1 (\mathcal{R}_1-1)+1)) (2 f_0 (\mathcal{H}_0 \mathcal{H}_2 r_2 (\mathcal{R}_2-1)+\mathcal{H}_0-\mathcal{H}_2)
           \right. \nonumber \\
    & \left. \qquad
    -3 \mathcal{H}_0 \Omega_{M0} (\mathcal{H}_2 r_2 (\mathcal{R}_2-1)+1)) \right\} \, ,
                      \\
    J_{00}^{(\chi)} &=
   -\frac{27 \chi_1^2 \chi_2^2 \mathcal{H}_0^4 }{4 \Delta \chi ^4 a(\chi_1) a(\chi_2)}
  \left(\mu ^2-1\right) \Omega_{M0}^2 D_1(\chi_1) (\chi_1-r_1) D_1(\chi_2) (\chi_2-r_2) 
               \nonumber \\
    &  \qquad
  \left(8 \mu  \left(\chi_1^2+\chi_2^2\right) -9 \chi_1 \chi_2 \mu ^2-7 \chi_1 \chi_2\right) \, ,
                          \\
    J_{02}^{(\chi)} &= 
    -\frac{27 \chi_1^2 \chi_2^2 \mathcal{H}_0^4}{2 \Delta \chi ^4 a(\chi_1) a(\chi_2)} \left(\mu ^2-1\right) \Omega_{M0}^2 D_1(\chi_1) (\chi_1-r_1) D_1(\chi_2) (\chi_2-r_2) 
      \nonumber \\
    &  \qquad
    \left(4 \mu  \left(\chi_1^2+\chi_2^2\right)-3 \chi_1 \chi_2 \mu ^2-5 \chi_1 \chi_2\right) \, ,
                              \\
    J_{40}^{(\chi)} &=
    \frac{81 \mathcal{H}_0^4 \Omega_{M0}^2 D_1(\chi_1) D_1(\chi_2)}{\mathcal{H}_1 \mathcal{H}_2 a(\chi_1) a(\chi_2)}
    \Delta \chi^4 \left[(f(\chi_1)-1) \mathcal{H}(\chi_1)
    (\mathcal{H}_1 r_1 (\mathcal{R}_1-1)+1)+\mathcal{H}_1\right]
            \nonumber \\
    &  \qquad  
    \left[(f(\chi_2)-1) \mathcal{H}(\chi_2) (\mathcal{H}_2 r_2 (\mathcal{R}_2-1)+1)+\mathcal{H}_2\right] \, ,
                                 \\
    J_{31}^{(\chi)} &= 
    -\frac{81 \mathcal{H}_0^4 \mu  \Omega_{M0}^2 D_1(\chi_1) D_1(\chi_2)}{\mathcal{H}_1 \mathcal{H}_2 a(\chi_1) a(\chi_2)}
   \Delta \chi^2
                \nonumber \\
    &  \qquad  
    \left\{\chi_1 \mathcal{H}_2 (f(\chi_1)-1) \mathcal{H}(\chi_1) (\chi_2-r_2) (\mathcal{H}_1 r_1 (\mathcal{R}_1-1)+1)
               \right. \nonumber \\
    & \left. \qquad
    +\chi_2 \mathcal{H}_1 (\chi_1-r_1) (f(\chi_2)-1) \mathcal{H}(\chi_2) (\mathcal{H}_2 r_2 (\mathcal{R}_2-1)+1)+\mathcal{H}_1 \mathcal{H}_2 (\chi_1 \chi_2-r_1 r_2)\right\} \, ,
                                 \\
    J_{22}^{(\chi)} &= 
\frac{81 \chi_1 \chi_2 \mathcal{H}_0^4 \Omega_{M0}^2 D_1(\chi_1) D_1(\chi_2) }{4 \Delta \chi ^4 \mathcal{H}_1 \mathcal{H}_2 a(\chi_1) a(\chi_2)}
\left\{\mathcal{H}_1 \mathcal{H}_2 \left[2 \chi_1^5 \left(\mu ^2 (5 \chi_2-6 r_2)-\chi_2+2 r_2\right)
           \right. \right. \nonumber \\
    & \left. \left. \qquad \quad
+2 \chi_1^4 \left(-16 \chi_2^2 \mu +2 \chi_2 \left(-3 \mu ^2 r_1+r_1+2 \left(\mu ^2+3\right) \mu  r_2\right)+\left(7 \mu ^2-3\right) r_1 r_2\right)
           \right. \right. \nonumber \\
    & \left. \left. \qquad \quad
-\chi_1^3 \chi_2 \left(\chi_2^2 \left(5 \mu ^4-22 \mu ^2-31\right)+\chi_2 \left(\mu ^2+3\right) \left(3 \left(\mu ^2+3\right) r_2-8 \mu  r_1\right)
           \right. \right. \right.\nonumber \\
    & \left. \left. \left. \qquad \qquad
+16 \mu  \left(\mu ^2+1\right) r_1 r_2\right)
           \right. \right. \nonumber \\
    & \left. \left. \qquad \quad
+\chi_1^2 \chi_2^2 \left(-32 \chi_2^2 \mu -\chi_2 \left(\mu ^2+3\right) \left(3 \left(\mu ^2+3\right) r_1-8 \mu  r_2\right)
           \right. \right. \right.\nonumber \\
    & \left. \left. \left. \qquad \qquad
+\left(11 \mu ^4+14 \mu ^2+23\right) r_1 r_2\right)
           \right. \right. \nonumber \\
    & \left. \left. \qquad \quad
+2 \chi_1 \chi_2^3 \left(\chi_2^2 \left(5 \mu ^2-1\right)+2 \chi_2 \left(2 \left(\mu ^2+3\right) \mu  r_1-3 \mu ^2 r_2+r_2\right)-8 \mu  \left(\mu ^2+1\right) r_1 r_2\right)
           \right. \right. \nonumber \\
    & \left. \left. \qquad \quad
+2 \chi_2^4 r_1 \left(\mu ^2 (7 r_2-6 \chi_2)+2 \chi_2-3 r_2\right)\right]
           \right.  \nonumber \\
    &  \left. \qquad
-2 \Delta \chi ^4 (\mu -1) (\mu +1) \left[\chi_1 \mathcal{H}_2 (f(\chi_1)-1) \mathcal{H}(\chi_1) (\chi_2-r_2) (\mathcal{H}_1 r_1 (\mathcal{R}_1-1)+1)
          \right. \right. \nonumber \\
    & \left. \left. \qquad \quad
+\chi_2 \mathcal{H}_1 (\chi_1-r_1) (f(\chi_2)-1) \mathcal{H}(\chi_2) (\mathcal{H}_2 r_2 (\mathcal{R}_2-1)+1) \right] \right\} \, ,
                                     \\
    J_{31}^{(\chi_1)} &= 
    -\frac{27 \chi_1^2 f_0 \mathcal{H}_0^3 \mu  \Omega_{M0} D_1(\chi_1)}{\mathcal{H}_1 \mathcal{H}_2 r_2 a(\chi_1)} (\mathcal{R}_2 \mathcal{H}_2 r_2  +1) (\chi_1 (f(\chi_1)-1) \mathcal{H}(\chi_1) 
   \nonumber \\
    &  \qquad   
    (\mathcal{H}_1 r_1 (\mathcal{R}_1-1)+1)+\mathcal{H}_1 r_1) \, ,
                                         \\
    J_{40}^{(\chi_1)} &= 
    \frac{27 \chi_1^4 \mathcal{H}_0^3 \Omega_{M0} D_1(\chi_1) }{2 \mathcal{H}_1 \mathcal{H}_2 r_2 a(\chi_1)}
    \left[(f(\chi_1)-1) \mathcal{H}(\chi_1) (\mathcal{H}_1 r_1 (\mathcal{R}_1-1)+1)+\mathcal{H}_1 \right]
      \nonumber \\
    &   \qquad   
    \left[2 f_0 (\mathcal{H}_0 \mathcal{H}_2 r_2 (\mathcal{R}_2-1)+\mathcal{H}_0-\mathcal{H}_2)-3 \mathcal{H}_0 \Omega_{M0} (\mathcal{H}_2 r_2 (\mathcal{R}_2-1)+1)\right] \, ,
                                             \\
    J_{00}^{(\Delta \chi_1)} &= 
    \frac{3 f_2 \mathcal{H}_0^2 \Omega_{M0} D_1(\chi_1)  }{10 \mathcal{H}_1 r_2 a(\chi_1)}
    \left\{
    \mathcal{H}_1 \left[2 \chi_1^3 \mathcal{R}_2 \mathcal{H}_2^2 \mu  r_2^2  
              \right.    \right.   \nonumber \\
    &   \left. \left. \qquad \quad 
    -\chi_1^2 \left(r_2 \left(\mathcal{R}_2 \mathcal{H}_2^2 r_2   \left(2 \mu  r_1+\left(\mu ^2+3\right) r_2\right)
               \right.     \right.    \right. \right.  \nonumber \\
    &   \left. \left. \left. \left. \qquad \qquad \quad
    -2 \mathcal{H}_2 (\mathcal{R}_2-1) \left(\mu  r_1+\left(2-6 \mu ^2\right) r_2\right)+12 \mu ^2-4\right)-2 \mu  r_1\right)
        \right.   \right.  \nonumber \\
    &  \left. \left. \qquad   \quad
    +\chi_1 r_2 \left(r_1 \left(\mathcal{H}_2 r_2 \left(\mathcal{R}_2   \left(\mathcal{H}_2 \left(\mu ^2+3\right) r_2+8 \mu ^2-6\right)-8 \mu ^2+6\right)+8 \mu ^2-6\right)
                 \right.     \right.    \right.   \nonumber \\
    &   \left. \left. \left. \qquad \qquad   
    +2 \mu  r_2 (\mathcal{H}_2 r_2 (\mathcal{R}_2   (\mathcal{H}_2 r_2+5)-5)+5)\right)
          \right.   \right.  \nonumber \\
    &  \left. \left. \qquad   \quad  
    -2 r_2^2 \left(r_2 \left(\mathcal{R}_2 \mathcal{H}_2^2 \mu  r_1 r_2  +\mathcal{H}_2 (\mathcal{R}_2-1) (2 \mu  r_1+r_2)+1\right)+2 \mu  r_1\right)\right]
       \right.  \nonumber \\
    &  \left. \qquad  
    -2 \Delta\chi_1^2 (f(\chi_1)-1) \mathcal{H}(\chi_1) (r_2-\chi_1 \mu ) (\mathcal{H}_1 r_1 (\mathcal{R}_1-1)+1) 
           \right.  \nonumber \\
    &  \left. \qquad  
    (\mathcal{H}_2 r_2 (\mathcal{R}_2-1)+1)
    \right\}
          \nonumber \\
    &     
    -\frac{3 \mathcal{H}_0^2 \Omega_{M0} D_1(\chi_1)}{20 \mathcal{H}_2 a(\chi_1) a_2} (\chi_1-r_1) 
               \nonumber \\
    &   \qquad  
    \left\{3 \mathcal{H}_0^2 \Omega_{M0} \left(2 \chi_1^2 \mu -\chi_1 \left(\mu ^2+3\right) r_2+2 \mu  r_2^2\right) (\mathcal{H}_2 r_2 (\mathcal{R}_2-2)+1)
               \right.  \nonumber \\
    &  \left. \qquad  
    -20 \mathcal{R}_2 \mathcal{H}_2 \mu  r_2 a_2\right\} \, ,
                                             \\
    J_{02}^{(\Delta \chi_1)} &= 
\frac{3 f_2 \mathcal{H}_0^2 \Omega_{M0} D_1(\chi_1)   }{14 \Delta\chi_1^2 \mathcal{H}_1 r_2 a(\chi_1)}
\nonumber \\
&
\left\{\mathcal{H}_1 \left[4 \chi_1^5 \mathcal{R}_2 \mathcal{H}_2^2 \mu  r_2^2  -2 \chi_1^4 \left(r_2 \left(\mathcal{R}_2 \mathcal{H}_2^2 r_2   \left(2 \mu  r_1+\left(5 \mu ^2+3\right) r_2\right)
\right. \right. \right. \right.
\nonumber \\
& \left. \left. \left. \left. \qquad \qquad \quad
-2 \mathcal{H}_2 (\mathcal{R}_2-1) \left(\mu  r_1+\left(2-6 \mu ^2\right) r_2\right)+12 \mu ^2-4\right) -2 \mu  r_1\right)
\right. \right.
\nonumber \\
& \left. \left. \qquad \quad
+\chi_1^3 r_2 \left(\mu  r_2 \left(4 \mathcal{R}_2 \mathcal{H}_2^2 \left(\mu ^2+5\right) r_2^2  
\right. \right. \right. \right.
\nonumber \\
& \left. \left. \left. \left. \qquad \qquad \quad
+13 \mathcal{H}_2 \left(3 \mu ^2+1\right) r_2 (\mathcal{R}_2-1)+39 \mu ^2+13\right)
\right. \right. \right.
\nonumber \\
& \left. \left. \left. \qquad \qquad
+2 r_1 \left(\mathcal{H}_2 r_2 \left(\mathcal{R}_2   \left(\mathcal{H}_2 \left(5 \mu ^2+3\right) r_2+4 \mu ^2-6\right)-4 \mu ^2+6\right)+4 \mu ^2-6\right)\right)
\right. \right.
\nonumber \\
& \left. \left. \qquad \quad
-\chi_1^2 r_2^2 \left(2 \mathcal{R}_2 \mathcal{H}_2^2 \left(5 \mu ^2+3\right) r_2^3  
\right. \right. \right.
\nonumber \\
& \left. \left. \left. \qquad \qquad
+\mathcal{H}_2 r_2^2 \left(\mathcal{R}_2   \left(\mu  \left(4 \mathcal{H}_2 \left(\mu ^2+5\right) r_1+55 \mu \right)+5\right)-55 \mu ^2-5\right)
\right. \right. \right.
\nonumber \\
& \left. \left. \left. \qquad \qquad
+r_2 \left(\mathcal{H}_2 \left(23 \mu ^2-11\right) \mu  r_1 (\mathcal{R}_2-1)+55 \mu ^2+5\right)+\mu  \left(23 \mu ^2-11\right) r_1\right)
\right. \right.
\nonumber \\
& \left. \left. \qquad \quad
+\chi_1 r_2^3 \left(r_1 \left(\mathcal{H}_2 r_2 \left(\mathcal{R}_2   \left(2 \mathcal{H}_2 \left(5 \mu ^2+3\right) r_2+23 \mu ^2-3\right)-23 \mu ^2+3\right)+23 \mu ^2-3\right)
\right. \right. \right.
\nonumber \\
& \left. \left. \left. \qquad \qquad
+4 \mu  r_2 (\mathcal{H}_2 r_2 (\mathcal{R}_2   (\mathcal{H}_2 r_2+7)-7)+7)\right)
\right. \right.
\nonumber \\
& \left. \left. \qquad \quad
-4 r_2^4 \left(r_2 \left(\mathcal{R}_2 \mathcal{H}_2^2 \mu  r_1 r_2  +\mathcal{H}_2 (\mathcal{R}_2-1) (2 \mu  r_1+r_2)+1\right)+2 \mu  r_1\right)\right]
\right.
\nonumber \\
& \left. \qquad 
+4 \Delta\chi_1^4 (f(\chi_1)-1) \mathcal{H}(\chi_1) (\chi_1 \mu -r_2) (\mathcal{H}_1 r_1 (\mathcal{R}_1-1)+1) 
\right.
\nonumber \\
& \left. \qquad 
(\mathcal{H}_2 r_2 (\mathcal{R}_2-1)+1)\right\}
               \nonumber \\
    &   
-\frac{3 \mathcal{H}_0^2 \Omega_{M0} D_1(\chi_1)}{14 \Delta\chi_1^2 \mathcal{H}_2 a(\chi_1) a_2} (\chi_1-r_1)   \left(2 \chi_1^2 \mu -\chi_1 \left(\mu ^2+3\right) r_2+2 \mu  r_2^2\right) 
               \nonumber \\
    &   \qquad
\left(3 \Delta\chi_1^2 \mathcal{H}_0^2 \Omega_{M0} (\mathcal{H}_2 r_2 (\mathcal{R}_2-2)+1)-7 \mathcal{R}_2 \mathcal{H}_2 r_2 a_2\right) \, ,
                                             \\
    J_{04}^{(\Delta \chi_1)} &= 
\frac{9 f_2 \mathcal{H}_0^2 \Omega_{M0} D_1(\chi_1)   }{70 \Delta\chi_1^2 \mathcal{H}_1 r_2 a(\chi_1)} 
\left\{\mathcal{H}_1 \left[2 \chi_1^5 \mathcal{R}_2 \mathcal{H}_2^2 \mu  r_2^2  
\right. \right.
\nonumber \\
& \left. \left. \qquad \quad
-\chi_1^4 \left(r_2 \left(\mathcal{R}_2 \mathcal{H}_2^2 r_2   \left(2 \mu  r_1+\left(5 \mu ^2+3\right) r_2\right)
\right. \right. \right. \right.
\nonumber \\
& \left. \left. \left. \left. \qquad \qquad \quad
-2 \mathcal{H}_2 (\mathcal{R}_2-1) \left(\mu  r_1+\left(2-6 \mu ^2\right) r_2\right)+12 \mu ^2-4\right)-2 \mu  r_1\right)
\right. \right.
\nonumber \\
& \left. \left. \qquad \quad
+\chi_1^3 r_2 \left(\mu  r_2 \left(2 \mathcal{R}_2 \mathcal{H}_2^2 \left(\mu ^2+5\right) r_2^2  
\right. \right. \right. \right.
\nonumber \\
& \left. \left. \left. \left. \qquad \qquad \quad
+\mathcal{H}_2 \left(9 \mu ^2+17\right) r_2 (\mathcal{R}_2-1)+9 \mu ^2+17\right)
\right. \right. \right.
\nonumber \\
& \left. \left. \left. \qquad \qquad
+r_1 \left(\mathcal{H}_2 r_2 \left(\mathcal{R}_2   \left(\mathcal{H}_2 \left(5 \mu ^2+3\right) r_2+4 \mu ^2-6\right)-4 \mu ^2+6\right)+4 \mu ^2-6\right)\right)
\right. \right.
\nonumber \\
& \left. \left. \qquad \quad
-\chi_1^2 r_2^2 \left(\mathcal{R}_2 \mathcal{H}_2^2 \left(5 \mu ^2+3\right) r_2^3  
\right. \right. \right.
\nonumber \\
& \left. \left. \left. \qquad \qquad
+\mathcal{H}_2 r_2^2 \left(\mathcal{R}_2   \left(\mu  \left(2 \mathcal{H}_2 \left(\mu ^2+5\right) r_1+17 \mu \right)+13\right)-17 \mu ^2-13\right)
\right. \right. \right.
\nonumber \\
& \left. \left. \left. \qquad \qquad
+r_2 \left(\mathcal{H}_2 \left(\mu ^2+5\right) \mu  r_1 (\mathcal{R}_2-1)+17 \mu ^2+13\right)+\mu  \left(\mu ^2+5\right) r_1\right)
\right. \right.
\nonumber \\
& \left. \left. \qquad \quad
+\chi_1 r_2^3 \left(r_1 \left(\mathcal{H}_2 r_2 \left(\mathcal{R}_2   \left(\mathcal{H}_2 \left(5 \mu ^2+3\right) r_2+\mu ^2+9\right)-\mu ^2-9\right)+\mu ^2+9\right)
\right. \right. \right.
\nonumber \\
& \left. \left. \left. \qquad \qquad
+2 \mu  r_2 (\mathcal{H}_2 r_2 (\mathcal{R}_2   (\mathcal{H}_2 r_2+7)-7)+7)\right)
\right. \right.
\nonumber \\
& \left. \left. \qquad \quad
-2 r_2^4 \left(r_2 \left(\mathcal{R}_2 \mathcal{H}_2^2 \mu  r_1 r_2  +\mathcal{H}_2 (\mathcal{R}_2-1) (2 \mu  r_1+r_2)+1\right)+2 \mu  r_1\right)\right]
\right.
\nonumber \\
& \left. \qquad 
+2 \Delta\chi_1^4 (f(\chi_1)-1) \mathcal{H}(\chi_1) (\chi_1 \mu -r_2) (\mathcal{H}_1 r_1 (\mathcal{R}_1-1)+1) 
\right.
\nonumber \\
& \left. \qquad 
(\mathcal{H}_2 r_2 (\mathcal{R}_2-1)+1)\right\}
               \nonumber \\
    &   
-\frac{27 \mathcal{H}_0^4 \Omega_{M0}^2 D_1(\chi_1) (\chi_1-r_1)   \left(2 \chi_1^2 \mu -\chi_1 \left(\mu ^2+3\right) r_2+2 \mu  r_2^2\right) (\mathcal{H}_2 r_2 (\mathcal{R}_2-2)+1)}{140 \mathcal{H}_2 a(\chi_1) a_2} \, ,
                                                 \\
    J_{20}^{(\Delta \chi_1)} &= 
    \frac{9 \Delta\chi_1^2 f_2 \mathcal{H}_0^2 \Omega_{M0} D_1(\chi_1)  }{\mathcal{H}_1 r_2 a(\chi_1)} 
    \nonumber \\
&  
    \left\{ 
    \mathcal{H}_1 
    \left(
    \mu  r_1 +
    r_2 \left(\mathcal{R}_2 \mathcal{H}_2^2 \mu  r_2   (\chi_1-r_1)-\mathcal{H}_2 (\mathcal{R}_2-1) (r_2-\mu  r_1)-1\right)
    \right)
    \right.
\nonumber \\
& \left. \qquad 
    -(f(\chi_1)-1) \mathcal{H}(\chi_1) (r_2-\chi_1 \mu ) (\mathcal{H}_1 r_1 (\mathcal{R}_1-1)+1) (\mathcal{H}_2 r_2 (\mathcal{R}_2-1)+1)\right\}
                   \nonumber \\
    &   
    -\frac{9 \Delta\chi_1^2 \mathcal{H}_0^2 \Omega_{M0} D_1(\chi_1)  }{2 \mathcal{H}_1 \mathcal{H}_2 a(\chi_1) a_2} \left\{2 \mathcal{R}_2 \mathcal{H}_2 a_2 ((f(\chi_1)-1) \mathcal{H}(\chi_1) (\mathcal{H}_1 r_1 (\mathcal{R}_1-1)+1)     +\mathcal{H}_1)
        \right.
\nonumber \\
& \left. \qquad 
+3 \mathcal{H}_0^2 \mathcal{H}_1 \mu  \Omega_{M0} (\chi_1-r_1) (\mathcal{H}_2 r_2 (\mathcal{R}_2-2)+1)\right\} \, ,
                                                     \\
    J_{40}^{(\Delta \chi_1)} &= 
    \frac{81 \Delta\chi_1^4 \mathcal{H}_0^4 \Omega_{M0}^2 D_1(\chi_1)}{2 \mathcal{H}_1 \mathcal{H}_2 r_2 a(\chi_1) a_2}
    (\mathcal{H}_2 r_2 (\mathcal{R}_2-2)+1) ((f(\chi_1)-1)
                   \nonumber \\
    &   
    \mathcal{H}(\chi_1) (\mathcal{H}_1 r_1 (\mathcal{R}_1-1)+1)+\mathcal{H}_1)
               \nonumber \\
    &   
    -\frac{27 \mathcal{R}_2 \Delta\chi_1^4 f_2 \mathcal{H}_0^2 \mathcal{H}_2^2 \Omega_{M0}}{\mathcal{H}_1 a(\chi_1)}   D_1(\chi_1)  
                   \nonumber \\
    & 
    \left[(f(\chi_1)-1) \mathcal{H}(\chi_1) (\mathcal{H}_1 r_1 (\mathcal{R}_1-1)+1)+\mathcal{H}_1 \right] \, ,
   \end{align}
where we have defined $\mathcal{R}= \bar \rho_M / \bar \rho_{\rm tot}$ and $f=\dot D_1/D_1/\mathcal{H}$ denotes the growth rate. The subscript $1$ and $2$ indicate that the quantity is evaluated at $r_1$ and $r_2$, respectively, while evaluation at comoving distance $\chi_1$ or $\chi_2$ is indicated. Note also that $0\leq \chi_1\leq r_1$ and $0\leq \chi_2\leq r_2$.

\section{Modified \class{} transfer functions}
\label{app:CLASS}
We modify the \class{} code~\cite{CLASS} to compute the angular power spectrum of the Hawking Energy. \class{} is already computing the galaxy number counts angular power spectrum~\cite{DiDio:2013bqa,DiDio:2016ykq} as
\begin{equation}
    C_\ell (z_1 , z_2) = \frac{2}{\pi} \int dk k^2 \Delta_\ell \left( k , z_1 \right) \Delta_\ell \left( k , z_2 \right) P_\mathcal{R} \left( k \right)
\end{equation}
where $P_{\mathcal{R}} \left( k \right) $ is the primordial curvature power spectrum and
\begin{equation}
     \Delta_\ell \left( k , z \right) = \sum_X      \Delta_\ell^{(X)} \left( k , z \right)
\end{equation}
is the full angular transfer function. We express the individual transfer functions in terms of the \class{} notation\footnote{The galaxy number counts transfer functions in \class{} are generalized for non-vanishing spatial curvature~\cite{DiDio:2016ykq}. Here we modify them directly, however in our derivation we have considered only a flat universe. Therefore, a priori, this implementation is correct for spatially flat FL universe only.}, ensuring that their implementation in the code is straightforward:
\begin{align}
    \Delta_\ell^{({\rm Den})_i} & = \int_0^{\eta_0} d\eta W_i \frac{\bar \rho_f}{\bar \rho_T} S_D \mathfrak{j}_{\ell}^{\kappa} \, ,\\
      \Delta_\ell^{({\rm Red})_i} & = 0 \, , \\
\Delta_\ell^{({\rm Len})_i} & = \ell \left( \ell +1 \right) \int_0^{\eta_0} d\eta W^L_i S_{\Phi + \Psi}\mathfrak{j}_{\ell}^{\kappa} \, , \\
\Delta_\ell^{({\rm D1})_i} & = \int_0^{\eta_0} d\eta W_i \left(
\frac{3 \left( 1 + \bar w \right) \frac{\bar \rho_f}{\bar \rho_T}-3}{k} +  \frac{3}{a H}\mathfrak{cot}_K(\chi)
\right) S_\Theta  {\mathfrak{j}_{\ell}^{\kappa}}' \, , \\
\Delta_\ell^{({\rm D2})_i} & = -3 \int_0^{\eta_0} d\eta W_i   \left( 1+ \bar w \right)  \frac{\bar \rho_f}{\bar \rho_T}   \frac{a H}{k^2} S_\Theta \mathfrak{j}_{\ell}^{\kappa}\, , \\
\Delta_\ell^{({\rm G1})_i} & = 0 \, , \\
\Delta_\ell^{({\rm G2})_i} & = - \int_0^{\eta_0} d\eta W_i\left( 
3 \left( 1 + \bar w \right) \frac{\bar \rho_f}{\bar \rho_T}  + \frac{3 k}{a H}  \mathfrak{cot}_K(\chi) 
\right)
S_\Phi \mathfrak{j}_{\ell}^{\kappa} \, , \\
\Delta_\ell^{({\rm G3})_i} & = 0 \,
 , \\
\Delta_\ell^{({\rm G4})_i} & = \int_0^{\eta_0} d\eta W_i^{\rm G4} S_{\Phi + \Psi} \mathfrak{j}_{\ell}^{\kappa} \,
 , \\
\Delta_\ell^{({\rm G5})_i} & = \int_0^{\eta_0} d\eta W_i^{\rm G5} S_{\Phi + \Psi} {\mathfrak{j}_{\ell}^{\kappa} }'\, ,
\end{align}
with the modified window functions defined by
\begin{align}
    W_i^L \left( \eta \right) & = - \frac{3}{2} \int_0^{\eta} d \tilde \eta W_i \left( \tilde \eta \right)  \frac{k\,\mathfrak{sin}_K(\chi-\tilde\chi)}{\mathfrak{sin}_K(\chi)\mathfrak{sin}_K(\tilde\chi)} \,  , \\
    W_i^{\rm G4} \left( \eta \right) & =
   3  \int_0^{\eta} d \tilde \eta W_i \left( \tilde \eta \right) 
     k \mathfrak{cot}_K(\tilde \chi) 
    \, , \\
    W_i^{\rm G5} \left( \eta \right) & = 
    \int_0^{\eta} d \tilde \eta W_i \left( \tilde \eta \right) 
    k\left( 3 \left( 1 + \bar w \right) \frac{\bar \rho_f}{\bar \rho_T}-3
         + \frac{3 k}{aH}\mathfrak{cot}_K(\tilde \chi) 
    \right)_{\tilde \eta} 
    \, , 
\end{align}
where $\bar{\rho}_f$ denotes the density of the cosmic fluid, whereas $\bar{\rho}_T=\bar{\rho}_f+\Lambda$ denotes the total density. These modifications are given in terms of a general cosmic fluid with pressure $\bar{P}_f=\bar{w}\,\bar{\rho}_f$. However, for the numerical results of Section \ref{sec:angularpowerspectra}, $\bar{w}=0$ because we study the case of dust plus a cosmological constant.

\bibliographystyle{JHEP}
\bibliography{refs}

\providecommand{\href}[2]{#2}\begingroup\raggedright\begin{thebibliography}{10}

\bibitem{Szabados:2009eka}
L.B.~Szabados, \emph{{Quasi-Local Energy-Momentum and Angular Momentum in
  General Relativity}}, \href{https://doi.org/10.12942/lrr-2009-4}{\emph{Living
  Rev. Rel.} {\bfseries 12} (2009) 4}.

\bibitem{Hawking:1968qt}
S.~Hawking, \emph{{Gravitational radiation in an expanding universe}},
  \href{https://doi.org/10.1063/1.1664615}{\emph{J. Math. Phys.} {\bfseries 9}
  (1968) 598}.

\bibitem{Stock:2020oda}
D.~Stock, \emph{{The Hawking Energy on the Past Lightcone in Cosmology}},
  \href{https://doi.org/10.1088/1361-6382/aba182}{\emph{Class. Quant. Grav.}
  {\bfseries 37} (2020) 215005}
  [\href{https://arxiv.org/abs/2003.13583}{{\ttfamily 2003.13583}}].

\bibitem{Stock:2020drm}
D.~Stock, \emph{{Applications of the Hawking Energy in Inhomogeneous
  Cosmology}}, \href{https://doi.org/10.1088/1361-6382/abe882}{\emph{Class.
  Quant. Grav.} {\bfseries 38} (2021) 075019}
  [\href{https://arxiv.org/abs/2010.07896}{{\ttfamily 2010.07896}}].

\bibitem{Durrer:2020fza}
R.~Durrer, \emph{{The Cosmic Microwave Background}}, Cambridge University Press
  (12, 2020),
  \href{https://doi.org/10.1017/9781316471524}{10.1017/9781316471524}.

\bibitem{Hayward:1993ph}
S.A.~Hayward, \emph{{Quasilocal gravitational energy}},
  \href{https://doi.org/10.1103/PhysRevD.49.831}{\emph{Phys. Rev. D} {\bfseries
  49} (1994) 831} [\href{https://arxiv.org/abs/gr-qc/9303030}{{\ttfamily
  gr-qc/9303030}}].

\bibitem{HorowitzSchmidt}
G.T.~Horowitz, B.G.~Schmidt and R.~Penrose, \emph{Note on gravitational
  energy}, \href{https://doi.org/10.1098/rspa.1982.0066}{\emph{Proceedings of
  the Royal Society of London. A. Mathematical and Physical Sciences}
  {\bfseries 381} (1982) 215}
  [\href{https://arxiv.org/abs/https://royalsocietypublishing.org/doi/pdf/10.1098/rspa.1982.0066}{{\ttfamily
  https://royalsocietypublishing.org/doi/pdf/10.1098/rspa.1982.0066}}].

\bibitem{Sasaki:1987ad}
M.~Sasaki, \emph{{The Magnitude - Redshift relation in a perturbed Friedmann
  universe}}, {\emph{Mon. Not. Roy. Astron. Soc.} {\bfseries 228} (1987) 653}.

\bibitem{Bonvin:2005ps}
C.~Bonvin, R.~Durrer and M.A.~Gasparini, \emph{{Fluctuations of the luminosity
  distance}}, \href{https://doi.org/10.1103/PhysRevD.85.029901}{\emph{Phys.
  Rev. D} {\bfseries 73} (2006) 023523}
  [\href{https://arxiv.org/abs/astro-ph/0511183}{{\ttfamily
  astro-ph/0511183}}].

\bibitem{Yoo:2016vne}
J.~Yoo and F.~Scaccabarozzi, \emph{{Unified Treatment of the Luminosity
  Distance in Cosmology}},
  \href{https://doi.org/10.1088/1475-7516/2016/09/046}{\emph{JCAP} {\bfseries
  09} (2016) 046} [\href{https://arxiv.org/abs/1606.08453}{{\ttfamily
  1606.08453}}].

\bibitem{Gasperini:2011us}
M.~Gasperini, G.~Marozzi, F.~Nugier and G.~Veneziano, \emph{{Light-cone
  averaging in cosmology: Formalism and applications}},
  \href{https://doi.org/10.1088/1475-7516/2011/07/008}{\emph{JCAP} {\bfseries
  07} (2011) 008} [\href{https://arxiv.org/abs/1104.1167}{{\ttfamily
  1104.1167}}].

\bibitem{Marozzi:2014kua}
G.~Marozzi, \emph{{The luminosity distance\textendash{}redshift relation up to
  second order in the Poisson gauge with anisotropic stress}},
  \href{https://doi.org/10.1088/0264-9381/32/4/045004}{\emph{Class. Quant.
  Grav.} {\bfseries 32} (2015) 045004}
  [\href{https://arxiv.org/abs/1406.1135}{{\ttfamily 1406.1135}}].

\bibitem{Scaccabarozzi:2018vux}
F.~Scaccabarozzi, J.~Yoo and S.G.~Biern, \emph{{Galaxy Two-Point Correlation
  Function in General Relativity}},
  \href{https://doi.org/10.1088/1475-7516/2018/10/024}{\emph{JCAP} {\bfseries
  10} (2018) 024} [\href{https://arxiv.org/abs/1807.09796}{{\ttfamily
  1807.09796}}].

\bibitem{Castorina:2021xzs}
E.~Castorina and E.~Di~Dio, \emph{{The observed galaxy power spectrum in
  General Relativity}},
  \href{https://doi.org/10.1088/1475-7516/2022/01/061}{\emph{JCAP} {\bfseries
  01} (2022) 061} [\href{https://arxiv.org/abs/2106.08857}{{\ttfamily
  2106.08857}}].

\bibitem{Planck}
{\scshape Planck} collaboration, \emph{{Planck 2018 results. I. Overview and
  the cosmological legacy of Planck}},
  \href{https://doi.org/10.1051/0004-6361/201833880}{\emph{Astron. Astrophys.}
  {\bfseries 641} (2020) A1}
  [\href{https://arxiv.org/abs/1807.06205}{{\ttfamily 1807.06205}}].

\bibitem{Anderson:2012sa}
L.~Anderson et~al., \emph{{The clustering of galaxies in the SDSS-III Baryon
  Oscillation Spectroscopic Survey: Baryon Acoustic Oscillations in the Data
  Release 9 Spectroscopic Galaxy Sample}},
  \href{https://doi.org/10.1111/j.1365-2966.2012.22066.x}{\emph{Mon. Not. Roy.
  Astron. Soc.} {\bfseries 427} (2013) 3435}
  [\href{https://arxiv.org/abs/1203.6594}{{\ttfamily 1203.6594}}].

\bibitem{DESI:2023bgx}
{\scshape DESI} collaboration, \emph{{First Detection of the BAO Signal from
  Early DESI Data}},  \href{https://arxiv.org/abs/2304.08427}{{\ttfamily
  2304.08427}}.

\bibitem{Brout:2022vxf}
D.~Brout et~al., \emph{{The Pantheon+ Analysis: Cosmological Constraints}},
  \href{https://doi.org/10.3847/1538-4357/ac8e04}{\emph{Astrophys. J.}
  {\bfseries 938} (2022) 110}
  [\href{https://arxiv.org/abs/2202.04077}{{\ttfamily 2202.04077}}].

\bibitem{Bonvin_2011}
C.~Bonvin and R.~Durrer, \emph{{What galaxy surveys really measure}},
  \href{https://doi.org/10.1103/PhysRevD.84.063505}{\emph{Phys. Rev. D}
  {\bfseries 84} (2011) 063505}
  [\href{https://arxiv.org/abs/1105.5280}{{\ttfamily 1105.5280}}].

\bibitem{Tutusaus:2022cab}
I.~Tutusaus, D.~Sobral-Blanco and C.~Bonvin, \emph{{Combining gravitational
  lensing and gravitational redshift to measure the anisotropic stress with
  future galaxy surveys}},  \href{https://arxiv.org/abs/2209.08987}{{\ttfamily
  2209.08987}}.

\bibitem{Lange:2021zre}
J.U.~Lange, A.P.~Hearin, A.~Leauthaud, F.C.~van~den Bosch, H.~Guo and
  J.~DeRose, \emph{{Five per\,cent measurements of the growth rate from
  simulation-based modelling of redshift-space clustering in BOSS LOWZ}},
  \href{https://doi.org/10.1093/mnras/stab3111}{\emph{Mon. Not. Roy. Astron.
  Soc.} {\bfseries 509} (2021) 1779}
  [\href{https://arxiv.org/abs/2101.12261}{{\ttfamily 2101.12261}}].

\bibitem{DESI:2016fyo}
{\scshape DESI} collaboration, \emph{{The DESI Experiment Part I:
  Science,Targeting, and Survey Design}},
  \href{https://arxiv.org/abs/1611.00036}{{\ttfamily 1611.00036}}.

\bibitem{Abate:2012za}
{LSST Dark Energy Science Collaboration}, \emph{{Large Synoptic Survey
  Telescope: Dark Energy Science Collaboration}}, {\emph{ArXiv e-prints} (2012)
  } [\href{https://arxiv.org/abs/1211.0310}{{\ttfamily 1211.0310}}].

\bibitem{Abell:2009aa}
{\scshape LSST Project} collaboration, \emph{{LSST Science Book, Version 2.0}},
   \href{https://arxiv.org/abs/0912.0201}{{\ttfamily 0912.0201}}.

\bibitem{Amendola:2016saw}
L.~Amendola et~al., \emph{{Cosmology and fundamental physics with the Euclid
  satellite}}, \href{https://doi.org/10.1007/s41114-017-0010-3}{\emph{Living
  Rev. Rel.} {\bfseries 21} (2018) 2}
  [\href{https://arxiv.org/abs/1606.00180}{{\ttfamily 1606.00180}}].

\bibitem{Laureijs:2011gra}
{\scshape Euclid} collaboration, \emph{{Euclid Definition Study Report}},
  \href{https://arxiv.org/abs/1110.3193}{{\ttfamily 1110.3193}}.

\bibitem{Maartens:2015mra}
{\scshape SKA Cosmology SWG} collaboration, \emph{{Overview of Cosmology with
  the SKA}}, {\emph{PoS} {\bfseries AASKA14} (2015) 016}
  [\href{https://arxiv.org/abs/1501.04076}{{\ttfamily 1501.04076}}].

\bibitem{CLASS}
J.~Lesgourgues, \emph{{The Cosmic Linear Anisotropy Solving System (CLASS) I:
  Overview}},  \href{https://arxiv.org/abs/1104.2932}{{\ttfamily 1104.2932}}.

\bibitem{DiDio:2013bqa}
E.~Di~Dio, F.~Montanari, J.~Lesgourgues and R.~Durrer, \emph{{The CLASSgal code
  for Relativistic Cosmological Large Scale Structure}},
  \href{https://doi.org/10.1088/1475-7516/2013/11/044}{\emph{JCAP} {\bfseries
  11} (2013) 044} [\href{https://arxiv.org/abs/1307.1459}{{\ttfamily
  1307.1459}}].

\bibitem{DiDio:2016ykq}
E.~Di~Dio, F.~Montanari, A.~Raccanelli, R.~Durrer, M.~Kamionkowski and
  J.~Lesgourgues, \emph{{Curvature constraints from Large Scale Structure}},
  \href{https://doi.org/10.1088/1475-7516/2016/06/013}{\emph{JCAP} {\bfseries
  06} (2016) 013} [\href{https://arxiv.org/abs/1603.09073}{{\ttfamily
  1603.09073}}].

\bibitem{Kogut:1993ag}
A.~Kogut et~al., \emph{{Dipole anisotropy in the COBE DMR first year sky
  maps}}, \href{https://doi.org/10.1086/173453}{\emph{Astrophys. J.} {\bfseries
  419} (1993) 1} [\href{https://arxiv.org/abs/astro-ph/9312056}{{\ttfamily
  astro-ph/9312056}}].

\bibitem{Lineweaver:1996xa}
C.H.~Lineweaver, L.~Tenorio, G.F.~Smoot, P.~Keegstra, A.J.~Banday and P.~Lubin,
  \emph{{The dipole observed in the COBE DMR four-year data}},
  \href{https://doi.org/10.1086/177846}{\emph{Astrophys. J.} {\bfseries 470}
  (1996) 38} [\href{https://arxiv.org/abs/astro-ph/9601151}{{\ttfamily
  astro-ph/9601151}}].

\bibitem{WMAP:2008ydk}
{\scshape WMAP} collaboration, \emph{{Five-Year Wilkinson Microwave Anisotropy
  Probe (WMAP) Observations: Data Processing, Sky Maps, and Basic Results}},
  \href{https://doi.org/10.1088/0067-0049/180/2/225}{\emph{Astrophys. J.
  Suppl.} {\bfseries 180} (2009) 225}
  [\href{https://arxiv.org/abs/0803.0732}{{\ttfamily 0803.0732}}].

\bibitem{Planck:2013kqc}
{\scshape Planck} collaboration, \emph{{Planck 2013 results. XXVII. Doppler
  boosting of the CMB: Eppur si muove}},
  \href{https://doi.org/10.1051/0004-6361/201321556}{\emph{Astron. Astrophys.}
  {\bfseries 571} (2014) A27}
  [\href{https://arxiv.org/abs/1303.5087}{{\ttfamily 1303.5087}}].

\bibitem{Planck:2020qil}
{\scshape Planck} collaboration, \emph{{Planck intermediate results. LVI.
  Detection of the CMB dipole through modulation of the thermal
  Sunyaev-Zeldovich effect: Eppur si muove II}},
  \href{https://doi.org/10.1051/0004-6361/202038053}{\emph{Astron. Astrophys.}
  {\bfseries 644} (2020) A100}
  [\href{https://arxiv.org/abs/2003.12646}{{\ttfamily 2003.12646}}].

\bibitem{Mitsou:2019ocs}
E.~Mitsou, J.~Yoo, R.~Durrer, F.~Scaccabarozzi and V.~Tansella, \emph{{General
  and consistent statistics for cosmological observations}},
  \href{https://doi.org/10.1103/PhysRevResearch.2.033004}{\emph{Phys. Rev.
  Res.} {\bfseries 2} (2020) 033004}
  [\href{https://arxiv.org/abs/1905.01293}{{\ttfamily 1905.01293}}].

\bibitem{Desjacques:2020zue}
V.~Desjacques, Y.B.~Ginat and R.~Reischke, \emph{{Statistics of a single sky:
  constrained random fields and the imprint of Bardeen potentials on galaxy
  clustering}},  \href{https://arxiv.org/abs/2009.02036}{{\ttfamily
  2009.02036}}.

\bibitem{Tansella:2018sld}
V.~Tansella, G.~Jelic-Cizmek, C.~Bonvin and R.~Durrer, \emph{{COFFE: a code for
  the full-sky relativistic galaxy correlation function}},
  \href{https://doi.org/10.1088/1475-7516/2018/10/032}{\emph{JCAP} {\bfseries
  10} (2018) 032} [\href{https://arxiv.org/abs/1806.11090}{{\ttfamily
  1806.11090}}].

\bibitem{Gourgoulhon:2005ng}
E.~Gourgoulhon and J.L.~Jaramillo, \emph{{A 3+1 perspective on null
  hypersurfaces and isolated horizons}},
  \href{https://doi.org/10.1016/j.physrep.2005.10.005}{\emph{Phys. Rept.}
  {\bfseries 423} (2006) 159}
  [\href{https://arxiv.org/abs/gr-qc/0503113}{{\ttfamily gr-qc/0503113}}].

\bibitem{Wald:1984rg}
R.M.~Wald, \emph{{General Relativity}}, Chicago Univ. Pr., Chicago, USA (1984),
  \href{https://doi.org/10.7208/chicago/9780226870373.001.0001}{10.7208/chicago/9780226870373.001.0001}.

\end{thebibliography}\endgroup

\end{document}